\documentclass{article}

\usepackage{arxiv}

\usepackage[utf8]{inputenc} 
\usepackage[T1]{fontenc}    
\usepackage{hyperref}       
\usepackage{url}            
\usepackage{booktabs}       
\usepackage{amsfonts}       
\usepackage{nicefrac}       
\usepackage{microtype}      
\usepackage{lipsum}		
\usepackage{graphicx}
\usepackage{natbib}
\usepackage{doi}
\usepackage{amsmath}
\usepackage{booktabs}
\usepackage{threeparttable}

\title{Foundation Model-Assisted Full Waveform Inversion}

\author{ \href{https://orcid.org/0009-0005-5239-3955}{\includegraphics[scale=0.06]{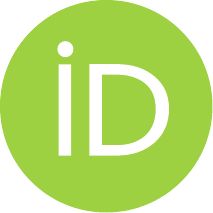}\hspace{1mm}Mustafa Alfarhan} \\
	KAUST\\
	Thuwal, Kingdom of Saudi Arabia\\
	\texttt{mustafa.alfarhan@kaust.edu.sa} \\
    \And
	{Matteo Ravasi} \\
	Shearwater GeoServices\\
	Gatwick, United Kingdom\\
	\texttt{mravasi@shearwatergeo.com} \\
	\And
	{Fuqiang Chen} \\
	KAUST\\
	Thuwal, Kingdom of Saudi Arabia\\
	\texttt{fuqiang.chen@kaust.edu.sa} \\
        \And
	{George Turkiyyah} \\
	KAUST\\
	Thuwal, Kingdom of Saudi Arabia\\
	\texttt{george.turkiyyah@kaust.edu.sa} \\
        \And
	{David Keyes} \\
	KAUST\\
	Thuwal, Kingdom of Saudi Arabia\\
	\texttt{david.keyes@kaust.edu.sa} \\
}

\renewcommand{\shorttitle}{Foundation Model-Assisted FWI}

\hypersetup{
hidelinks,
pdftitle={Foundation Model-Assisted Full Waveform Inversion},
pdfsubject={FWI},
pdfauthor={Mustafa Alfarhan, Matteo Ravasi, Fuqiang Chen, George Turkiyyah, and David Keyes},
pdfkeywords={Full waveform inversion, foundation model, neural networks, automatic differentiation, adjoint sources},
}

\begin{document}
\maketitle

\begin{abstract}
Full waveform inversion (FWI) can recover high-resolution subsurface velocity models. Conventional waveform-difference objectives, however, are vulnerable to cycle skipping when the starting model is inaccurate. We introduce an FWI objective that compares features produced from modeled and observed seismic traces by SeisLM, a pretrained seismic foundation model. The SeisLM encoder remains frozen during inversion, and the feature discrepancy is differentiated with respect to the modeled traces to construct an adjoint source compatible with the standard adjoint-state framework. We also test a scheduled hybrid loss that combines the SeisLM feature loss with the conventional $L_2$ objective. Time-shift diagnostics show that the loss computed from features produced by the pretrained encoder has a broader and smoother basin around the correct alignment than either the waveform $L_2$ objective or the feature loss obtained from an encoder with the same architecture and randomly initialized parameters. In the Marmousi experiment, the SeisLM and hybrid objectives produce similar improvements during early-stage inversion and provide useful models for subsequent reflection-based $L_2$ refinement. In the 2D Overthrust experiment, which begins from a laterally invariant linear-gradient model, the SeisLM feature-loss workflow outperforms the conventional and hybrid workflows, indicating that introducing the $L_2$ contribution too early can reintroduce cycle-skipping sensitivity. In the 3D Overthrust experiment, conventional $L_2$ inversion stalls near the initial linear gradient, whereas the SeisLM feature loss guides the inversion toward a background model from which $L_2$ refinement recovers the principal structures. These results support using features produced by pretrained seismic networks to define early-stage FWI objectives rather than complete replacements for waveform-domain misfits.
\end{abstract}


\section{Introduction}
Full waveform inversion (FWI) estimates subsurface properties by minimizing the discrepancy between observed and modeled seismic data. When the starting model is sufficiently accurate, the conventional $L_2$ waveform-difference objective can exploit amplitude and phase information to recover high-resolution velocity updates \cite{tarantola1984inversion, virieux2009overview}. However, comparing oscillatory seismic traces sample-by-sample makes this objective vulnerable to cycle skipping: if the predicted and observed arrivals differ by more than roughly half a period, the inversion may converge to a spurious local minimum rather than to the true model.
 
A broad range of strategies has been developed to enlarge the basin of attraction of FWI. One line of work modifies the misfit function itself. Cross-correlation and matching-filter objectives emphasize traveltime shifts rather than sample-by-sample amplitude differences \cite{luo1991wave, van2010correlation, warner2016adaptive}, while envelope, instantaneous-phase, optimal-transport, and dynamic-time-warping formulations reduce the oscillatory character of the residual or explicitly promote temporal alignment \cite{Bozdag2011845, chi2014full, wu2014seismic, engquist2016optimal, metivier2016measuring, yang2018application, ramos2019long, ma2013wave, chen2022cycle, kalita2023soft}. Comparative studies suggest that objectives emphasizing kinematic information are particularly useful during the early stages of inversion \cite{pladys2021cycle}. Other approaches expand the inverse problem, for example, through model extension, which relaxes the wave-equation constraint so that data fitting and physical consistency can be balanced during model updating \cite{vanleeuwen2013mitigating, aghamiry2019improving}. Acquisition and bandwidth also play a central role: low frequencies, long offsets, and wide apertures provide the long-wavelength information needed to recover the background velocity model \cite{sirgue2004efficient, sirgue2006importance, virieux2009overview, wang2019salt, cobo2019full, ji2025acquisition}. Although these approaches have been highly influential, they can require assumptions about the data, careful parameter tuning, changes to acquisition design, or information that is unavailable in legacy surveys. This motivates data-driven objectives that learn waveform similarity measures directly from seismic data.
 
Several data-driven misfit functions have already been proposed for FWI. In a meta-learning framework, \cite{sun2022mlmisfit} train a neural network to act directly as the FWI misfit, updating its parameters by running inversions on randomly generated velocity models and minimizing the resulting model error; the learned data-adaptive misfit has been shown to mitigate cycle skipping. A related line of work uses adversarial training, in which a discriminator network supplies the data-comparison signal: \cite{yang2023fwigan} employ a Wasserstein generative adversarial network with gradient penalty as a self-supervised loss during inversion, and \cite{wang2026fftinversiongan} extend this idea to the frequency domain. SiameseFWI \cite{saad2024siamesefwi} uses two weight-sharing CNN branches to map observed and simulated data into a shared latent space, where their feature distance defines a self-supervised FWI misfit that requires no pretraining or labeled data. F-SiameseFWI \cite{saad2025f} extends this idea to frequency-domain inputs and has been shown to reduce crosstalk in multi-source FWI while accelerating convergence. These approaches share the goal of replacing the handcrafted $L_2$ comparison with a learned one. However, each requires a network to be trained on an FWI-related objective, either offline through inversion-in-the-loop meta-learning or online through adversarial optimization performed during inversion.
 
This training requirement reflects a broader limitation of many deep-learning applications in seismic processing and interpretation: models are often task-specific and trained for a particular output, dataset, or geological setting \cite{wu2019inversionnet, yang2019deep, kumar2022deep, liu2022deep}. Foundation models instead learn reusable representations from broad data collections before being applied to downstream tasks. In machine learning, models such as BERT \cite{devlin2019bert}, GPT-2 \cite{radford2019language}, Wav2Vec2 \cite{baevski2020wav2vec}, HuBERT \cite{hsu2021hubert}, and masked autoencoders \cite{he2022masked} have demonstrated the value of self-supervised pretraining for learning transferable representations from large unlabeled datasets.
 
Related self-supervised pretraining approaches have also been explored in seismology. StorSeismic applies BERT-style pretraining to shot gathers for exploration-geophysics tasks \cite{harsuko2022storseismic}, while seismic foundation models learn from large collections of seismic images or migrated cubes for interpretation tasks \cite{sheng2023seismic, ordonez2026ncs}. In earthquake seismology, SeisCLIP and SeisT use multimodal or supervised pretraining for monitoring tasks \cite{si2024seisclip, li2024seist}. The Seismic Language Model (SeisLM) differs in that it is trained directly on unlabeled three-component seismic waveforms using a Wav2Vec2-like masked prediction objective \cite{liu2024seislm}. After pretraining, its encoder produces contextualized waveform features that summarize the temporal structure of seismic traces without being tied to a single downstream task.
 
Here, we investigate whether learned waveform representations provide a more robust space in which to compare modeled and observed FWI data. Instead of training a new FWI-specific misfit, we repurpose a frozen, self-supervised seismic foundation model as a fixed feature extractor. The objective is therefore defined by a network pretrained on unrelated earthquake waveforms to produce the representations compared during inversion, which lets us test whether seismic features, learned without any reference to FWI, can improve the robustness of early-stage FWI. The idea is conceptually related to perceptual losses in computer vision, where images are compared in the feature space of a pretrained neural network rather than pixel by pixel \cite{johnson2016perceptual, gatys2016image, dosovitskiy2016generating, simonyan2015very}. Analogously, comparing modeled and observed seismograms in the latent space of a deep neural network pretrained on seismological waveforms may emphasize structural waveform similarity while reducing sensitivity to small sample-wise phase errors.
 
We propose a SeisLM-based objective that measures the discrepancy between modeled and observed traces using the last-layer representations of a frozen SeisLM encoder (i.e., with fixed parameters). No training or fine-tuning is performed during inversion. Because the encoder remains differentiable with respect to its input, the feature-loss gradient can be computed by backpropagation and used as an adjoint source in a standard adjoint-state FWI workflow. Numerical examples show that this objective improves early-stage inversion by producing models that are less affected by cycle skipping than those obtained with the standard $L_2$ norm. These results indicate that the SeisLM feature loss can provide a more kinematically robust early-stage objective, guiding inversion from inaccurate starting models toward a better basin of attraction before conventional waveform-based refinement.
\section{Theoretical Framework}
This section first describes the SeisLM encoder, which will be used here as a feature extractor. It then defines the SeisLM feature loss and the hybrid loss used in FWI, and finally shows how the corresponding adjoint sources are computed.

\subsection{SeisLM}
\label{sec:seislm}

The proposed method uses SeisLM, a seismic foundation model pretrained through self-supervised learning \cite{liu2024seislm}. Only the components required for the FWI objective are summarized here; see \cite{liu2024seislm} for the complete architecture.

SeisLM follows the masked-prediction framework used by Wav2Vec2 \cite{baevski2020wav2vec} and BERT \cite{devlin2019bert}. A shallow 1D convolutional encoder maps a three-component waveform to a compressed sequence of latent vectors of dimension $d_v$,
\begin{equation}
    \mathbf{x} \in \mathbb{R}^{3 \times T}
    \;\longrightarrow\;
    \mathbf{v} = (\mathbf{v}_1,\ldots,\mathbf{v}_L),
    \qquad \mathbf{v}_{\ell} \in \mathbb{R}^{d_v}, \quad L<T .
\end{equation}
The resulting convolutional features are augmented with a Wav2Vec2-style relative positional embedding, computed using a grouped 1D convolution with a kernel size of 128 and 16 groups, before being passed through the transformer blocks to produce contextualized representations. Because this positional encoding does not use a fixed absolute position table, it can accommodate input lengths different from those used during pretraining. During pretraining, selected latent vectors are masked, and the transformer is trained with a contrastive loss to distinguish the correct quantized target from negative samples, i.e., quantized targets drawn from other time steps. This objective encourages the model to capture temporal waveform structure without requiring event labels, phase picks, or source information.

\begin{figure}[!ht]
\centering
\includegraphics[width=0.7\textwidth]{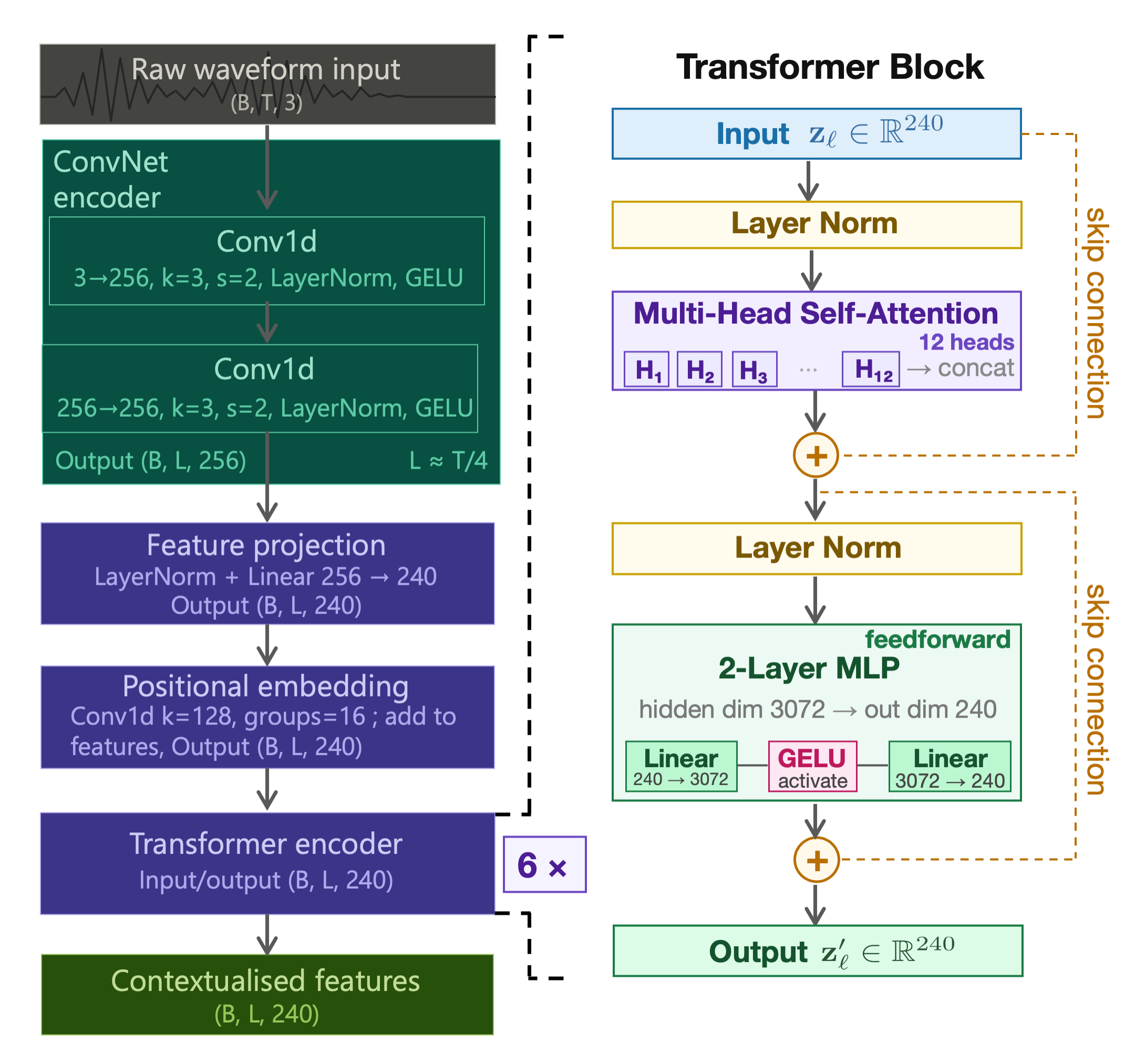}
\caption{SeisLM-base encoder used as a frozen feature extractor. Two stride-2 convolutional layers reduce the input length to \(L\approx T/4\), and six transformer blocks produce 240-dimensional contextualized features.}
\label{fig:seislm}
\end{figure}

SeisLM was pretrained on earthquake recordings from eight open-source datasets provided through SeisBench \cite{woollam2022seisbench}. We use SeisLM-base, which contains six transformer blocks and 11.4 million parameters. During inversion, we follow the downstream use of SeisLM and retain only the feature-extraction path: the convolutional encoder, positional embedding, and transformer layers. The masking module and quantization branch, which are used only to define the self-supervised pretraining task, are omitted during inversion, as shown in Figure~\ref{fig:seislm}.

For a source--receiver pair $(s,r)$, let $\mathbf{d}_{s,r}\in\mathbb{R}^{T}$ denote a single-component trace. Before applying SeisLM, each trace is standardized independently over time,
\[
    \mathcal{N}(\mathbf{d}_{s,r})
    =
    \frac{\mathbf{d}_{s,r}-\mu(\mathbf{d}_{s,r})}
    {\max(\sigma(\mathbf{d}_{s,r}), \epsilon)},
\]
where $\mu(\cdot)$ and $\sigma(\cdot)$ denote the mean and standard deviation, respectively, computed over the time samples of each trace, and $\epsilon>0$ is a small constant used for numerical stability. Because SeisLM expects a three-component input, define the replication
operator $\mathcal{R}:\mathbb{R}^{T}\rightarrow\mathbb{R}^{3\times T}$ by
\[
\mathcal{R}\mathbf{z}
=
\begin{bmatrix}
\mathbf{z}^{\top}\\
\mathbf{z}^{\top}\\
\mathbf{z}^{\top}
\end{bmatrix}.
\]
The network input is therefore
\[
\tilde{\mathbf{d}}_{s,r}
=
\mathcal{R}\mathcal{N}(\mathbf{d}_{s,r})
\in\mathbb{R}^{3\times T}.
\]
Under channel-stacked vectorization, $\mathcal{R}$ is represented by
\[
\begin{bmatrix}
\mathbf{I}_T\\
\mathbf{I}_T\\
\mathbf{I}_T
\end{bmatrix}.
\]
For a three-channel gradient
\[
\mathbf{G}
=
\begin{bmatrix}
\mathbf{g}_1^{\top}\\
\mathbf{g}_2^{\top}\\
\mathbf{g}_3^{\top}
\end{bmatrix}
\in\mathbb{R}^{3\times T},
\]
the adjoint replication operator satisfies
\[
\mathcal{R}^{\top}\mathbf{G}
=
\mathbf{g}_1+\mathbf{g}_2+\mathbf{g}_3,
\]
and therefore sums the gradients from the three input channels. The frozen encoder is written as
\begin{equation}\label{eq:seislm_forward}
    F_{\hat{\theta}}\!\left(\tilde{\mathbf{d}}_{s,r}\right)
    =
    \mathbf{A}_{s,r}
    =
    \begin{bmatrix}
    \mathbf{a}_1^{\top}\\
    \mathbf{a}_2^{\top}\\
    \vdots\\
    \mathbf{a}_{L}^{\top}
    \end{bmatrix}
    \in \mathbb{R}^{L\times 240},
\end{equation}
where $\hat{\theta}$ denotes the fixed network parameters, $L\approx T/4$, and each $\mathbf{a}_{\ell}\in\mathbb{R}^{240}$ is a contextualized temporal feature vector (the transformer embedding dimension; the convolutional latent dimension is $d_v=256$). SeisLM was pretrained on 30-s waveforms sampled at 100 Hz. In our experiments, the modeled and observed traces are retained on the native FWI time grid, introducing differences in both sampling rate and input duration. The numerical examples test whether the pretrained representations remain useful under this domain shift.

\subsection{SeisLM feature loss}
\label{sec:feature_loss}

\begin{figure}[!ht]
\centering
\includegraphics[width=\textwidth]{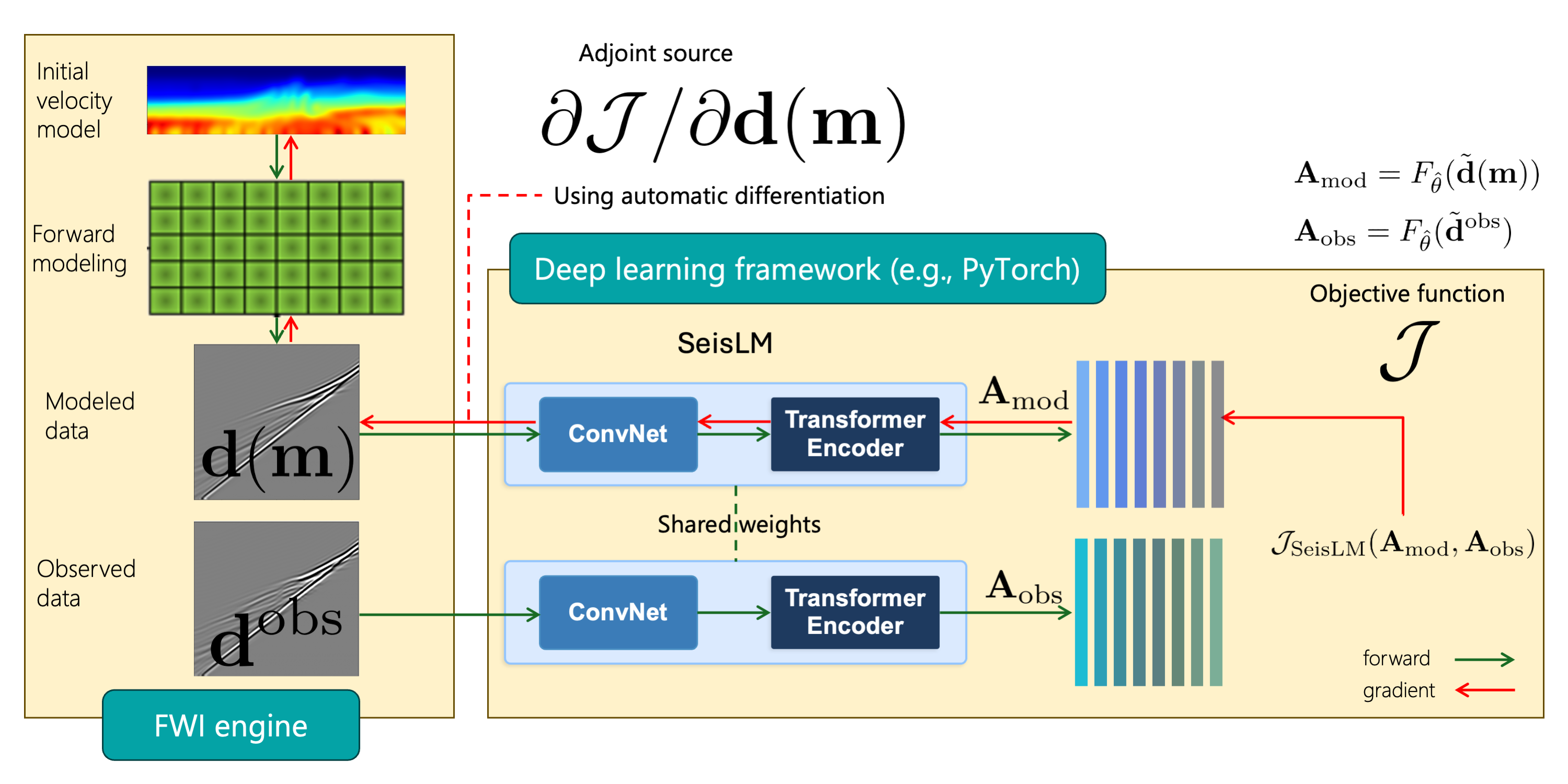}
\caption{FWI workflow using SeisLM as a frozen feature-space objective. The feature discrepancy defines \(\mathcal{J}_{\rm SeisLM}\), and backpropagation through the encoder provides the adjoint source used in the FWI update.}
\label{fig:seislm_fwi}
\end{figure}

The feature loss compares modeled and observed traces in the representation space of the frozen encoder. Traces from each shot gather are processed independently, with receivers batched for computational efficiency. Before entering the encoder, the modeled and observed traces undergo the preprocessing used to train SeisLM: each trace is standardized independently over time by subtracting its mean and dividing by $\max(\sigma,\epsilon)$. Because the acoustic trace is copied into three identical channels, the input differs from the physically distinct three-component earthquake recordings used during pretraining. Replication is the natural symmetric choice and ensures that all three channels contain waveform data, as they did during pretraining. Other mappings are possible, such as placing the trace in one channel and setting the remaining channels to zero, however these mappings introduce a different input-distribution shift. The experiments therefore test whether SeisLM's learned temporal representations transfer to acoustic FWI without relying on the original three-component interpretation. The feature loss is defined as
\begin{equation}\label{eq:feature_loss}
    \mathcal{J}_{\text{SeisLM}}(\mathbf{m})
    =
    \frac{1}{2}
    \sum_s \sum_r
    \left\|
    F_{\hat{\theta}}\!\left(\tilde{\mathbf{d}}_{s,r}(\mathbf{m})\right)
    -
    F_{\hat{\theta}}\!\left(\tilde{\mathbf{d}}_{s,r}^{\text{obs}}\right)
    \right\|_F^2 ,
\end{equation}
where 
\[
\tilde{\mathbf{d}}_{s,r}(\mathbf{m})
=
\mathcal{R}\mathcal{N}\left(\mathbf{d}_{s,r}(\mathbf{m})\right),
\qquad
\tilde{\mathbf{d}}_{s,r}^{\text{obs}}
=
\mathcal{R}\mathcal{N}\left(\mathbf{d}_{s,r}^{\text{obs}}\right).
\]
The modeled data are obtained by solving
$\mathcal{L}(\mathbf{m})\mathbf{u}_s=\mathbf{q}_s$ and sampling the
wavefield at the receiver locations:
\[
\mathbf{d}_s(\mathbf{m})=\mathbf{P}\mathbf{u}_s.
\]
For the observed traces, $\mathcal{N}$ can be applied once when caching the observed SeisLM features. For the modeled traces, however, $\mathcal{N}$ is part of the objective-function chain, so the adjoint source includes the derivative of the per-trace standardization as well as the derivative of the frozen SeisLM encoder. The modeled traces depend on the velocity model $\mathbf{m}$ through the forward wave equation. Observed features are computed and stored before inversion, whereas modeled features are recomputed at each iteration. The resulting workflow is illustrated in Figure~\ref{fig:seislm_fwi}.

\subsection{Hybrid loss}
\label{sec:hybrid_loss}

The SeisLM feature loss is intended to emphasize structural waveform similarity, whereas the conventional $L_2$ loss preserves direct sensitivity to waveform amplitudes and fine-scale details. As such, we propose to combine the two objectives as
\begin{equation}\label{eq:hybrid_loss}
    \mathcal{J}_{\text{hybrid}}(\mathbf{m})
    =
    \alpha_i \, \bar{\mathcal{J}}_{L_2}(\mathbf{m})
    +
    (1-\alpha_i) \, \bar{\mathcal{J}}_{\text{SeisLM}}(\mathbf{m}),
\end{equation}
where
\begin{equation}\label{eq:l2_loss}
    \mathcal{J}_{L_2}(\mathbf{m})
    =
    \frac{1}{2}
    \sum_s \sum_r
    \left\|
    \mathbf{d}_{s,r}(\mathbf{m})
    -
    \mathbf{d}_{s,r}^{\text{obs}}
    \right\|_2^2 .
\end{equation}
The normalized objectives are defined by $\bar{\mathcal{J}}_{q}(\mathbf{m})=\mathcal{J}_{q}(\mathbf{m})/\mathcal{J}_{q}(\mathbf{m}_0)$, where $q \in \{L_2, \mathrm{SeisLM}\}$ and $\mathbf{m}_0$ is the initial model. Normalization balances the initial magnitudes of the two terms, while the iteration-dependent weighting provides a gradual transition to the \(L_2\) waveform misfit. The iteration-dependent parameter $\alpha_i\in[0,1]$ controls the relative contributions of the two objectives. In the hybrid experiment, a linear schedule is used,
\begin{equation}\label{eq:alpha_schedule}
    \alpha_i = \min\left(1,\frac{i}{100}\right),
\end{equation}
where $i$ is the outer iteration index. Thus, the inversion starts with $\alpha_0=0$, corresponding to a pure SeisLM feature loss, and the $L_2$ contribution increases linearly until $\alpha_i=1$ at iteration 100. After this transition, the objective is equivalent to the conventional $L_2$ loss up to the constant normalization factor $\mathcal{J}_{L_2}(\mathbf{m}_0)^{-1}$.

\subsection{Adjoint source via backpropagation}
\label{sec:adjoint}
For an objective function that depends on the modeled data, the adjoint-state method gives the model gradient as (with $\mathbf{u}_s$ held fixed in the partial derivative)
\begin{equation}\label{eq:deriv_J_m_seislm}
\nabla_{\mathbf{m}}\mathcal{J}
=
-\sum_s
\left(
\frac{\partial[\mathcal{L}(\mathbf{m})\mathbf{u}_s]}
     {\partial\mathbf{m}}
\right)^{\top}
\boldsymbol{\lambda}_s,
\end{equation}
where the adjoint state $\boldsymbol{\lambda}_s$ satisfies
\begin{equation}\label{eq:adj_eq_seislm}
\mathcal{L}(\mathbf{m})^{\top}\boldsymbol{\lambda}_s
=
\frac{\partial\mathcal{J}}{\partial\mathbf{u}_s}
=
\mathbf{P}^{\top}
\frac{\partial\mathcal{J}}{\partial\mathbf{d}_s}.
\end{equation}
Here, $\mathcal{L}(\mathbf{m})^{\top}$ is the adjoint wave operator, and
$\mathbf{P}^{\top}\partial\mathcal{J}/\partial\mathbf{d}_s$ maps the
data-space adjoint source to the wavefield grid at the receiver locations.

For the $L_2$ loss, the adjoint source reduces to the data residual. For the SeisLM loss, the data-space adjoint source is obtained by backpropagating the feature residual through the frozen encoder, channel-replication operator, and per-trace standardization:
\begin{equation}\label{eq:adjoint_source_seislm}
\begin{aligned}
\frac{\partial \mathcal{J}_{\mathrm{SeisLM}}}
     {\partial \mathbf{d}_{s,r}}
&=
\left(
\frac{\partial \mathcal{N}(\mathbf{d}_{s,r})}
     {\partial \mathbf{d}_{s,r}}
\right)^{\top}
\mathcal{R}^{\top}
\left(
\frac{\partial F_{\hat{\theta}}(\tilde{\mathbf{d}}_{s,r})}
     {\partial \tilde{\mathbf{d}}_{s,r}}
\right)^{\top} \\
&\quad\times
\left[
F_{\hat{\theta}}(\tilde{\mathbf{d}}_{s,r})
-
F_{\hat{\theta}}(\tilde{\mathbf{d}}_{s,r}^{\mathrm{obs}})
\right].
\end{aligned}
\end{equation}
For notational simplicity, the Jacobians in Eq.~\ref{eq:adjoint_source_seislm} are understood to act on vectorized forms of the matrix-valued quantities. The adjoint replication operator $\mathcal{R}^{\top}$ sums the gradients from the three replicated input channels, while the Jacobian of $\mathcal{N}$ accounts for the dependence of the trace mean and standard deviation on the modeled data. Both operations are included automatically when the loss is differentiated through the preprocessing and frozen encoder.

The hybrid adjoint source is the corresponding weighted sum,
\begin{equation}\label{eq:hybrid_adjoint}
    \frac{\partial \mathcal{J}_{\text{hybrid}}}{\partial \mathbf{d}_{s,r}}
    =
    \frac{\alpha_i}{\mathcal{J}_{L_2}(\mathbf{m}_0)}
    \left(
    \mathbf{d}_{s,r}(\mathbf{m})
    -
    \mathbf{d}_{s,r}^{\text{obs}}
    \right)
    +
    \frac{1-\alpha_i}{\mathcal{J}_{\text{SeisLM}}(\mathbf{m}_0)}
    \frac{\partial \mathcal{J}_{\text{SeisLM}}}{\partial \mathbf{d}_{s,r}} .
\end{equation}
This source is injected into the adjoint wave equation in the same manner as the conventional residual. The resulting model gradient is supplied to the L-BFGS optimizer \cite{liu1989limited}, yielding the update
\begin{equation}\label{eq:model_update_seislm}
    \mathbf{m}_{i+1}
    =
    \mathbf{m}_i
    +
    \beta_i \mathbf{p}_i,
\end{equation}
where $\mathbf{p}_i$ is the L-BFGS search direction and $\beta_i$ is determined by line search.

\section{Numerical Examples}
\subsection{1D Example: Sensitivity to Time Shifts}
\label{subsec:toy_1d_sensitivity}

Before applying the proposed feature-space objective to FWI, its sensitivity to controlled time shifts is examined in a simple toy 1D setting. This experiment is not intended to represent a complete seismic acquisition geometry. Rather, it serves as a diagnostic illustration of the local shape of the objective function when seismic events are shifted in time, which is the basic mechanism underlying cycle skipping.

The observed trace is constructed as the sum of two Ricker wavelets with dominant frequency 3~Hz located at fixed arrival times. The modeled trace is then generated by independently shifting the two events by $\Delta t_1$ and $\Delta t_2$. For each pair of shifts, three normalized misfit values are computed: the conventional $L_2$ misfit in the data space, the feature-space $L_2$ misfit obtained from the last-layer hidden states of the pretrained SeisLM encoder, and the same feature-space misfit obtained using the same SeisLM encoder with randomly initialized parameters. The shift range spans approximately $\pm 2.5$ dominant periods for each event.

\begin{figure}[!ht]
\centering
\includegraphics[width=\textwidth]{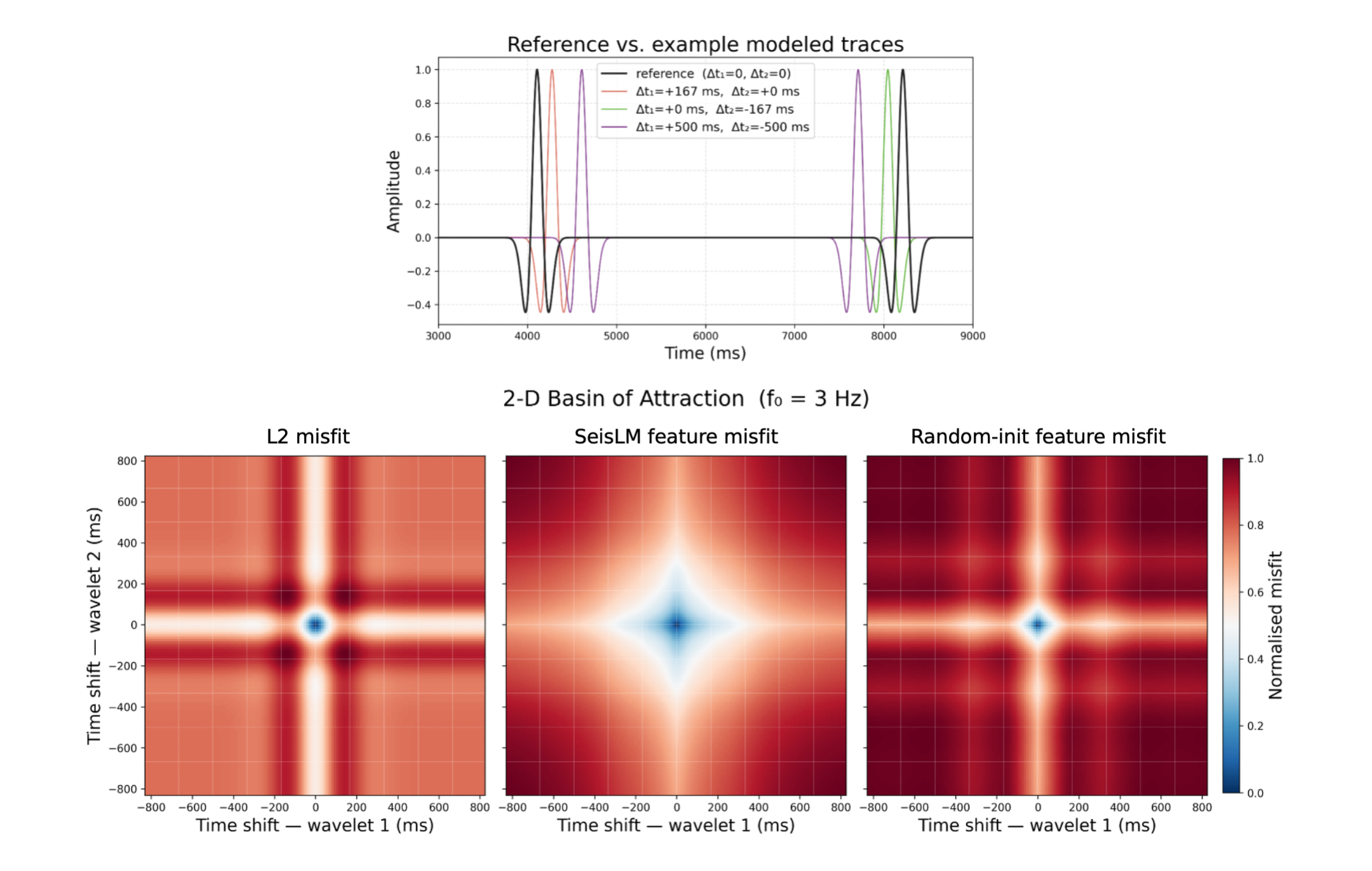}
\caption{1D example showing sensitivity to two independent time shifts. Top: reference and shifted two-wavelet traces. Bottom: normalized misfit surfaces for the raw $L_2$ norm, the pretrained SeisLM feature-space misfit, and the randomly initialized feature-space misfit. The pretrained feature space produces a broader and smoother basin around the correct solution.}
\label{fig:seislm_boa}
\end{figure}

Figure~\ref{fig:seislm_boa} shows the resulting two-dimensional misfit landscapes. The raw $L_2$ objective exhibits narrow low-misfit valleys aligned with the individual event shifts, reflecting the oscillatory sensitivity of waveform matching to phase errors. The randomly initialized network tested here does not consistently remove this structure, indicating that architecture alone does not explain the smoother landscape produced by the pretrained encoder. In contrast, the pretrained SeisLM feature-space misfit produces a broader and smoother basin around the correct solution $(\Delta t_1,\Delta t_2)=(0,0)$. This suggests that the learned representation is less sensitive to small event-wise phase shifts and may therefore provide a more robust objective during the early stages of FWI, when the modeled and observed arrivals are not yet accurately aligned.

\begin{figure}[!ht]
\centering
\includegraphics[width=\textwidth]{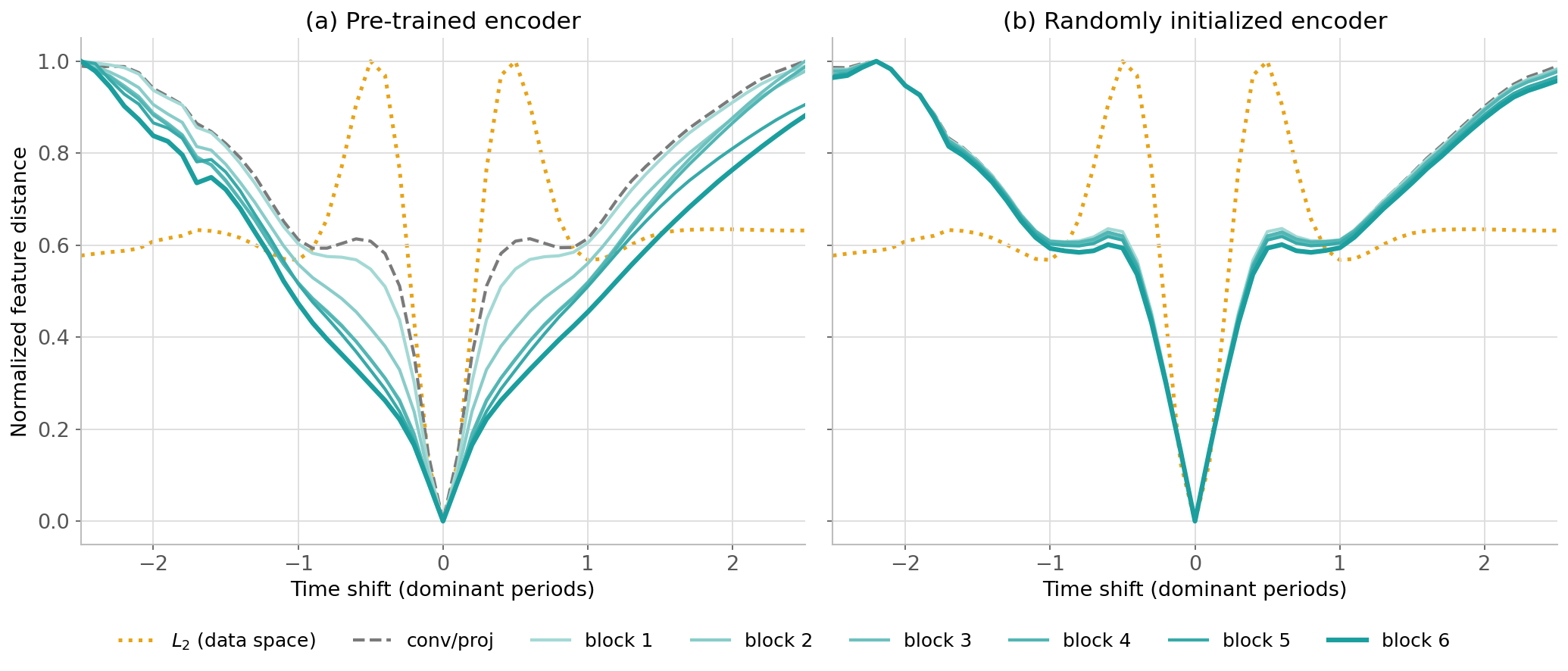}
\caption{Sensitivity of the SeisLM feature distance to a rigid time shift, resolved by encoder depth, for (a) the pretrained and (b) a randomly initialized encoder of the same architecture, with the data-space $L_2$ distance (dotted) for reference. Shifts are in dominant periods; each curve is normalized by its own maximum.}
\label{fig:layerwise}
\end{figure}

The 1D example isolates the role of pretraining in the final feature space, however it does not indicate where within the encoder that behavior originates. A second diagnostic addresses this question using the synthetic observed Marmousi data. A single shot gather, after applying the same early-arrival mask used in the refraction stage, is shifted rigidly in time by up to 2.5 dominant periods, and the distance between the shifted and unshifted gather is evaluated separately at the convolutional projection and at the output of each of the six transformer blocks. Figure~\ref{fig:layerwise} shows the resulting sensitivity profiles for the pretrained encoder and for a randomly initialized encoder of identical architecture, together with the conventional $L_2$ distance measured in the data space. Each curve is normalized by its own maximum, so the comparison concerns the shape of the profile rather than its magnitude.

In the pretrained encoder, the data-space $L_2$ distance increases only up to half a dominant period and then turns over, producing the oscillatory profile that underlies cycle skipping, and the convolutional features inherit this behavior. The transformer blocks, in contrast, increase monotonically over a substantially wider range of shifts, and the ramp becomes more gradual with depth, so the deeper representations retain a usable slope over a wider range of timing errors. This supports the use of last-layer features in Eq.~\ref{eq:feature_loss}. The randomly initialized encoder shows none of this structure: every layer turns over near the same shift as the $L_2$ distance, and the curves for different depths are nearly indistinguishable. Depth alone therefore does not produce the widened, monotonic profile, which emerges only in combination with pretraining. Repeating the diagnostic on the synthetic traces of the one-dimensional example gives the same ordering.

Following these diagnostic tests, the feature-space objective is evaluated using two benchmark velocity models. The Marmousi example provides a complex 2D test in which a smoothed starting model is updated using the conventional $L_2$ objective, the SeisLM feature loss, and the hybrid loss before reflection-based $L_2$ refinement. The SEG/EAGE Overthrust model is then used in 2D and 3D experiments. In both Overthrust experiments, the starting model is a laterally invariant linear velocity gradient $v(z)$ that contains no lateral structural information. The time--space mask used in stage 1 is constructed from the computed direct-arrival time for each source--receiver pair and suppresses later arrivals. The Marmousi and Overthrust experiments use Ricker wavelets with dominant frequencies of 7 and 5~Hz, respectively. Within each experiment, all objectives are compared using the same source wavelet. These tests examine whether the feature-space objective can recover a useful background model from increasingly complex models and acquisition geometries.

For plotting, each loss curve is normalized by its value at the beginning of the corresponding stage. Therefore, loss magnitudes are not directly comparable across objectives; only the evolution of each curve should be interpreted, whereas the relative velocity-error curves provide the cross-objective comparison of model accuracy. The normalized-loss curves include all objective-function evaluations performed by L-BFGS, including trial steps evaluated during line search. In contrast, the velocity-error curves are evaluated only for accepted model iterates, so they track the sequence of models retained by the optimizer rather than every line-search trial. Consequently, the loss curves should be interpreted as optimization histories rather than as a strict count of accepted model updates, and their lengths may differ between objectives or stages. An inversion is terminated when the L-BFGS convergence criteria are satisfied or when the line search cannot identify an acceptable step that satisfies the descent conditions.

All acoustic modeling and adjoint simulations are implemented with Devito~\cite{devito-api, devito-compiler}, and the L-BFGS optimization is carried out using SciPy~\cite{2020SciPy-NMeth}.

\subsection{Marmousi}
\begin{figure}[!ht] 
\centering
\includegraphics[width=\textwidth]{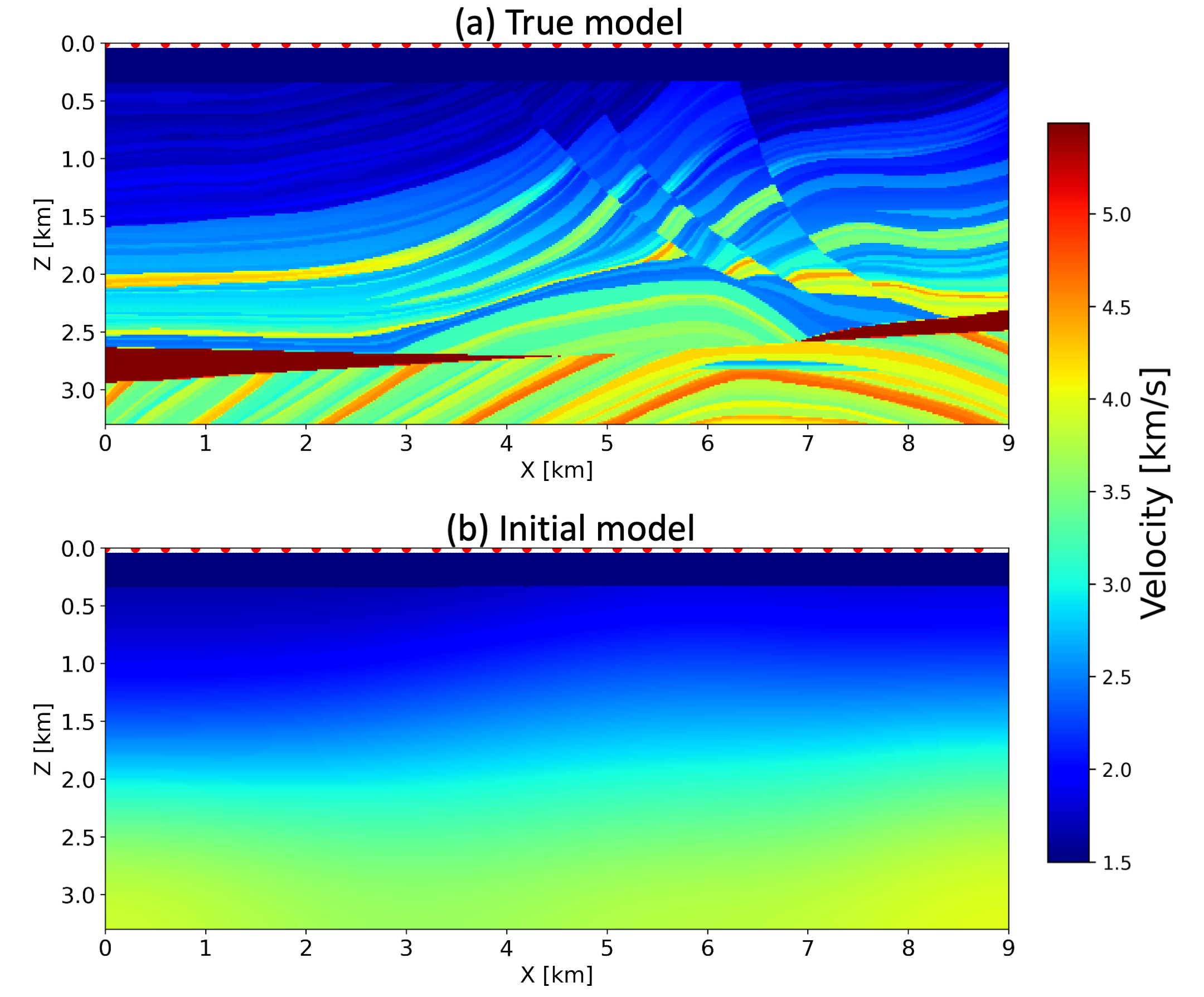} 
\caption{The Marmousi model: (a) true velocity model and (b) smoothed initial velocity model, with sources and receivers shown in red and white, respectively.}
\label{fig:marm_4} 
\end{figure}

Because of its complex geological structure, the synthetic Marmousi model~\cite{brougois1990marmousi} is widely used to evaluate seismic imaging and inversion methods. In this example, the model contains 601 grid points in the horizontal direction and 221 grid points in depth, with a uniform grid spacing of 15~m. Figure~\ref{fig:marm_4} shows the true velocity model and its smoothed counterpart, which serves as the initial model for FWI. The initial model is highly smoothed and sufficiently inaccurate to induce cycle skipping. The observed data are simulated with a variable-velocity, constant-density acoustic finite-difference modeling engine using 30 sources and 300 receivers evenly distributed at zero depth. The source signature is a Ricker wavelet with a peak frequency of 7~Hz.

The inversion is performed in two stages. In stage 1, a time-space mask suppresses reflection energy so that the update is driven primarily by refractions, using the conventional $L_2$ objective, the SeisLM feature loss, or a hybrid loss that combines the two. In stage 2, each corresponding stage-1 model is refined with the complete data set using the conventional $L_2$ objective. This separation allows us to test whether the feature-space objective provides a more kinematically accurate model for subsequent waveform-based refinement.

\begin{figure}[!ht] 
\centering
\includegraphics[width=\textwidth]{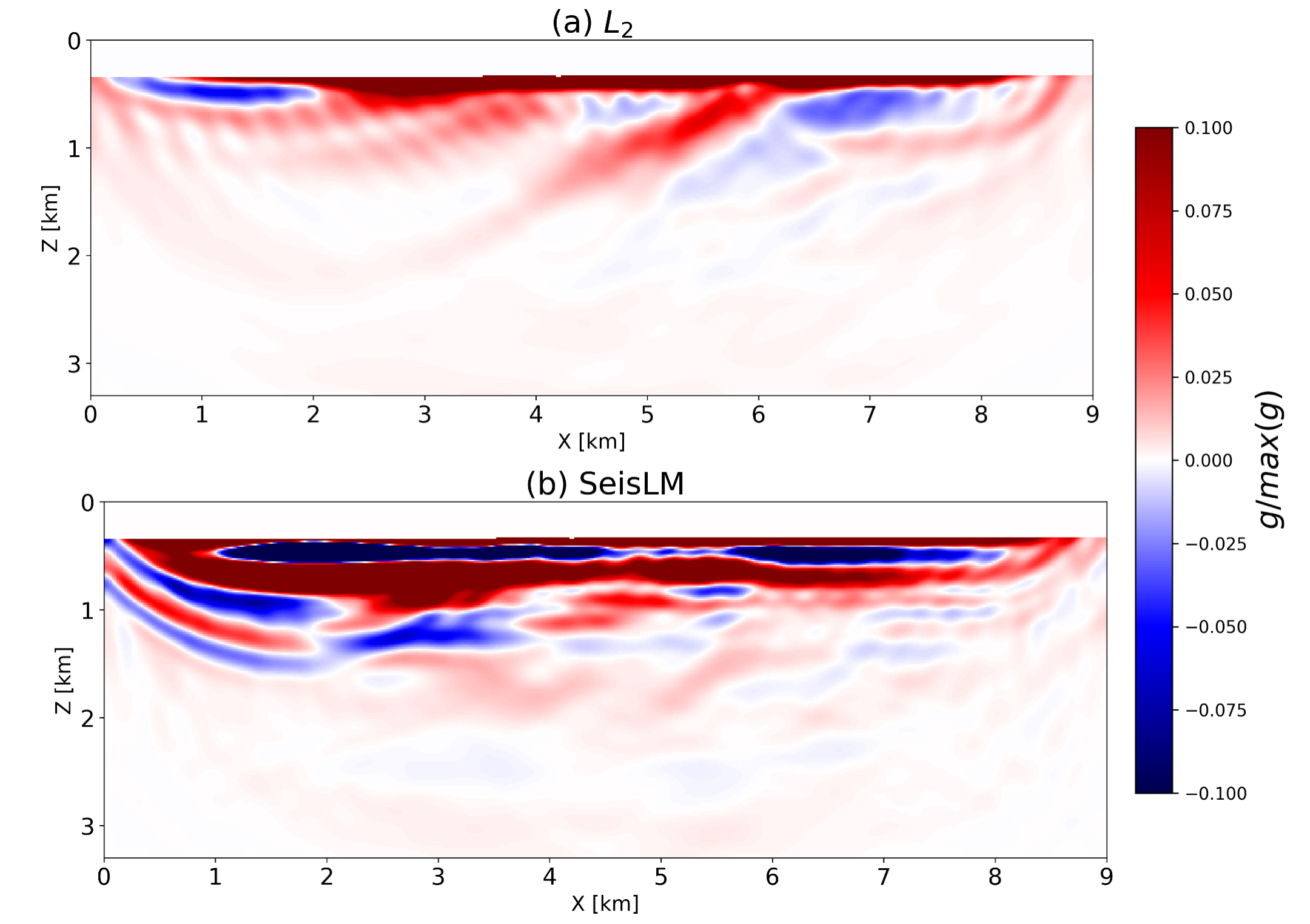} 
\caption{First-iteration FWI gradients obtained using (a) the conventional $L_2$ objective and (b) the SeisLM feature loss.}
\label{fig:marm_grad1} 
\end{figure}

Figures \ref{fig:marm_grad1}(a) and (b) show the first-iteration FWI gradients computed using the conventional $L_2$ objective and the SeisLM feature loss, respectively. The $L_2$ gradient is smooth and diffuse, with significant energy concentrated near the surface and gradually diminishing with depth. In contrast, the SeisLM gradient exhibits sharper, higher-amplitude updates that extend deeper into the model, with more laterally coherent structures that more closely resemble the structural complexity of the target model. This suggests that the feature loss provides a richer gradient for updating the background model during stage 1.

\begin{figure}[!ht] 
\centering
\includegraphics[width=\textwidth]{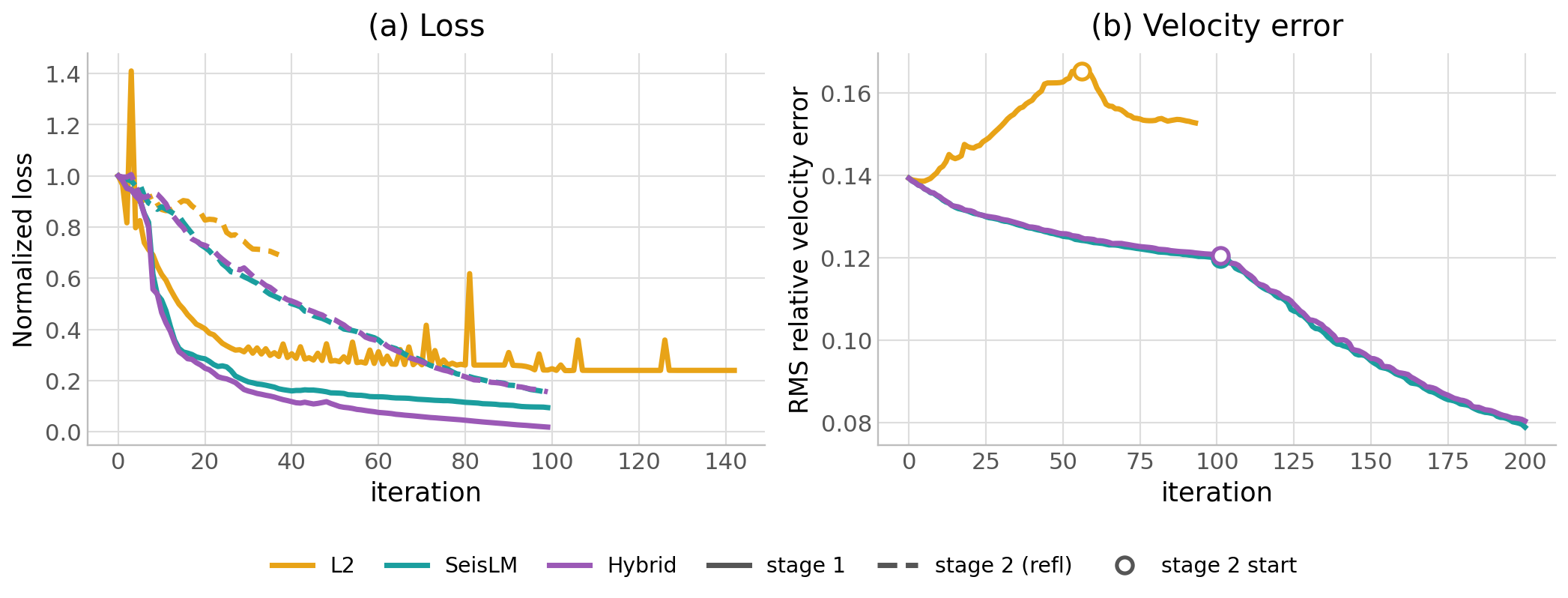} 
\caption{Performance of the Marmousi example. (a) Normalized loss and (b) relative velocity error. Solid curves denote stage 1, dashed curves denote reflection-based $L_2$ refinement, and circles mark the start of the reflection stage.}
\label{fig:marm_perf} 
\end{figure}

The convergence curves in Figure~\ref{fig:marm_perf} show how these gradient differences affect the inversion. During stage 1, the $L_2$ objective decreases while the relative velocity error increases, indicating convergence toward an inaccurate background model. In contrast, the SeisLM and hybrid objectives reduce the velocity error and follow similar convergence trajectories, providing more accurate starting models for the reflection stage. During stage 2, reflection-based $L_2$ refinement further reduces the loss and velocity error for both the SeisLM- and hybrid-initialized workflows. The $L_2$-initialized workflow begins the reflection stage from a less accurate background model and remains at a higher velocity error.

\begin{figure}[!ht] 
\centering
\includegraphics[width=\textwidth]{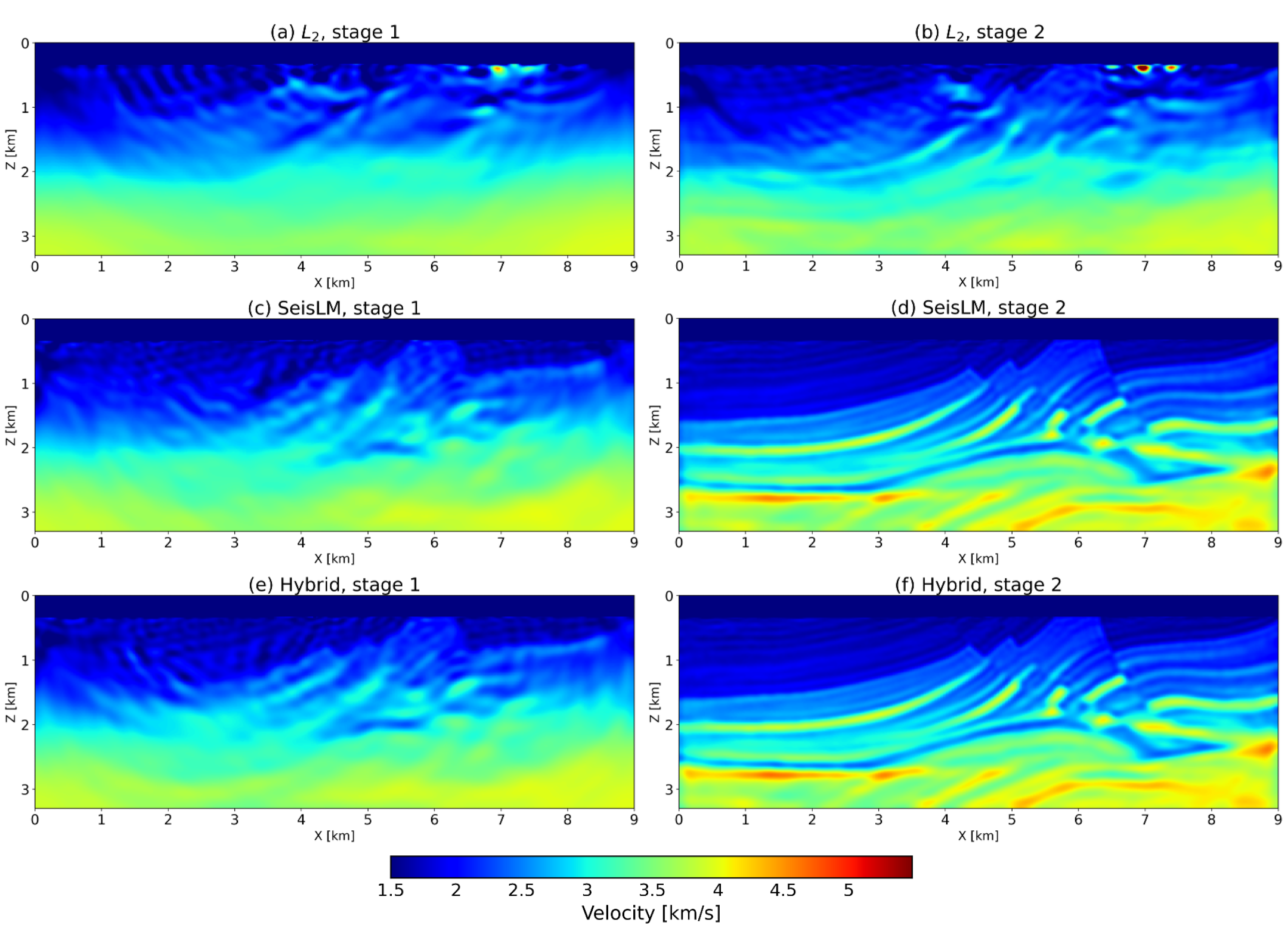} 
\caption{Inverted velocity models for the Marmousi example. Left column: stage-1 results obtained using (a) the conventional $L_2$ objective, (c) the SeisLM feature loss, and (e) the hybrid loss. Right column: corresponding models after reflection-based $L_2$ refinement for (b) $L_2$, (d) SeisLM, and (f) the hybrid workflow.}
\label{fig:marm_inv} 
\end{figure}

Figure~\ref{fig:marm_inv} compares the corresponding inverted velocity models at the end of stage 1 and after reflection-based $L_2$ refinement. Consistent with the velocity-error curves, the stage-1 $L_2$ result is strongly contaminated by artifacts, particularly in the shallow part of the model. In contrast, the SeisLM and hybrid stage-1 results recover more geologically coherent background models, with clearer shallow layering and smoother lateral continuity at depth. After reflection refinement, the $L_2$-initialized result remains affected by shallow artifacts and does not recover the main Marmousi structures reliably. The SeisLM- and hybrid-initialized results recover the principal dipping reflectors, the high-velocity zone at depth, and the complex fault structure in the central portion of the model more clearly, with the SeisLM and hybrid workflows providing similarly accurate final reconstructions.
 

\subsection{The SEG/EAGE Overthrust Model}
The SEG/EAGE Overthrust model~\cite{aminzadeh1997three} is a widely used 3D synthetic benchmark for seismic imaging and inversion. It contains strong lateral velocity contrasts, thrust faults, and steeply dipping reflectors. The original model is defined on an $801 \times 801 \times 187$ grid with a spacing of 25~m. For the 2D and 3D experiments, a 500~m water column with a velocity of 1.5~km/s is added above the selected model. The water layer is held fixed during inversion. Below the water layer, both experiments start from a laterally invariant, linearly increasing velocity profile $v(z)$ that contains no information about the true lateral structure. A 5~Hz Ricker wavelet is used in both experiments.

The Overthrust inversions are performed in two stages. In stage 1, a time--space mask suppresses reflection energy so that the background update is controlled mainly by refractions. The 2D experiment compares the conventional $L_2$, SeisLM, and hybrid objectives, whereas the 3D experiment compares only the $L_2$ and SeisLM objectives. In stage 2, the resulting stage-1 models are refined using the complete data set and the conventional $L_2$ objective. This separation allows us to test whether the feature loss provides a more kinematically accurate model for subsequent waveform-based refinement.

\subsubsection{2D Surface Acquisition}
\begin{figure}[!ht]
\centering
\includegraphics[width=\textwidth]{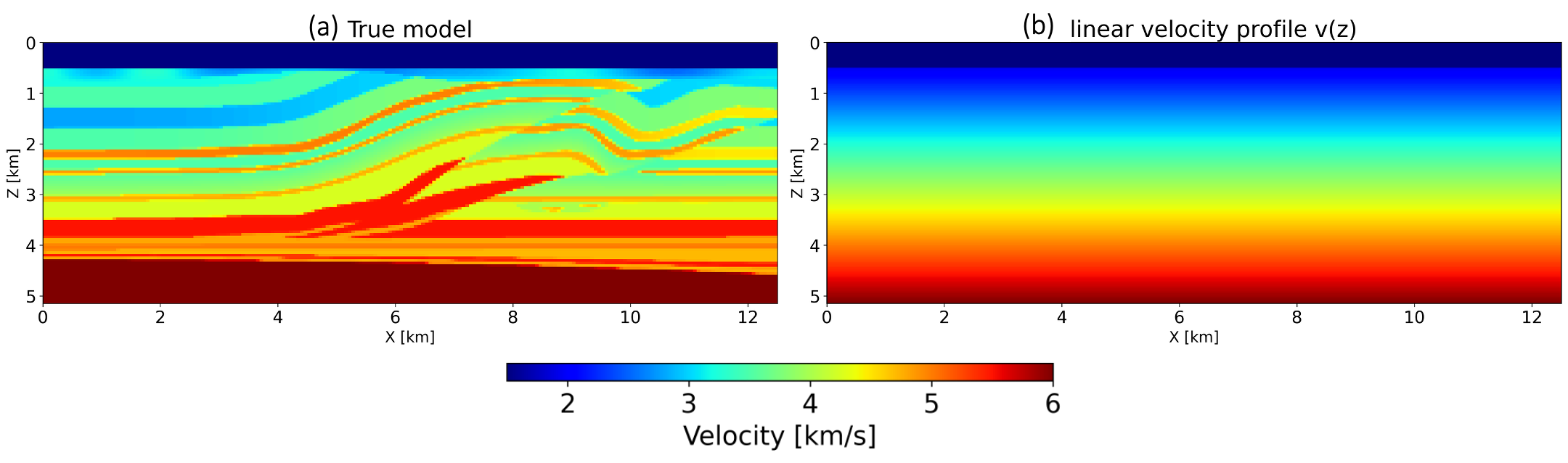}
\caption{The 2D Overthrust model with a 500~m water column: (a) true velocity model and (b) laterally invariant linear-gradient initial model.}
\label{fig:ovth_2d_model}
\end{figure}

In this work, a representative 2D section corresponding to the 456th inline slice along the $y$-axis ($y \approx 11.375$~km) is extracted and truncated to a lateral extent of 12.5~km, yielding a $501 \times 187$ model sampled at 25~m in both directions, with a physical extent of $12.5 \times 4.65$~km before the water column is added. After adding the 500~m water column, the total depth is approximately 5.2~km. A 2D fixed-spread surface acquisition geometry is employed, consisting of 30 uniformly distributed sources and 501 receivers spanning the full lateral extent of the model. Sources are spaced 430~m apart, with the first source at $x=0$~km and the last at $x=12.47$~km. Receivers are co-located with the surface grid nodes at a uniform spacing of 25~m. Both sources and receivers are positioned at the water surface ($z=0$). Each shot is recorded for 4~s. Figure~\ref{fig:ovth_2d_model} shows the true model and the laterally invariant linear-gradient initial model.

\begin{figure}[!ht]
\centering
\includegraphics[width=\textwidth]{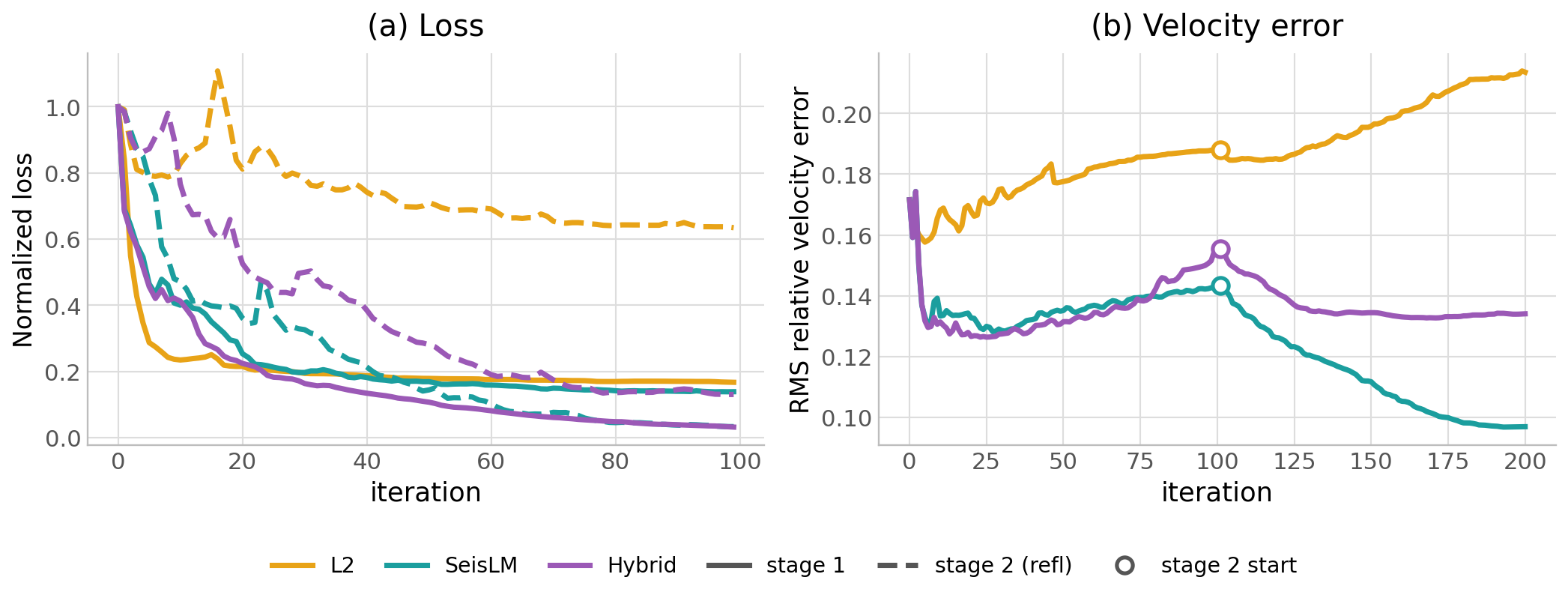}
\caption{Performance of the 2D Overthrust experiment. (a) Normalized objective values and (b) relative velocity error. Solid curves denote stage 1, dashed curves denote stage-2 $L_2$ refinement, and circles mark the start of stage 2.}
\label{fig:ovth_lininit_perf}
\end{figure}

\begin{figure}[!ht]
\centering
\includegraphics[width=\textwidth]{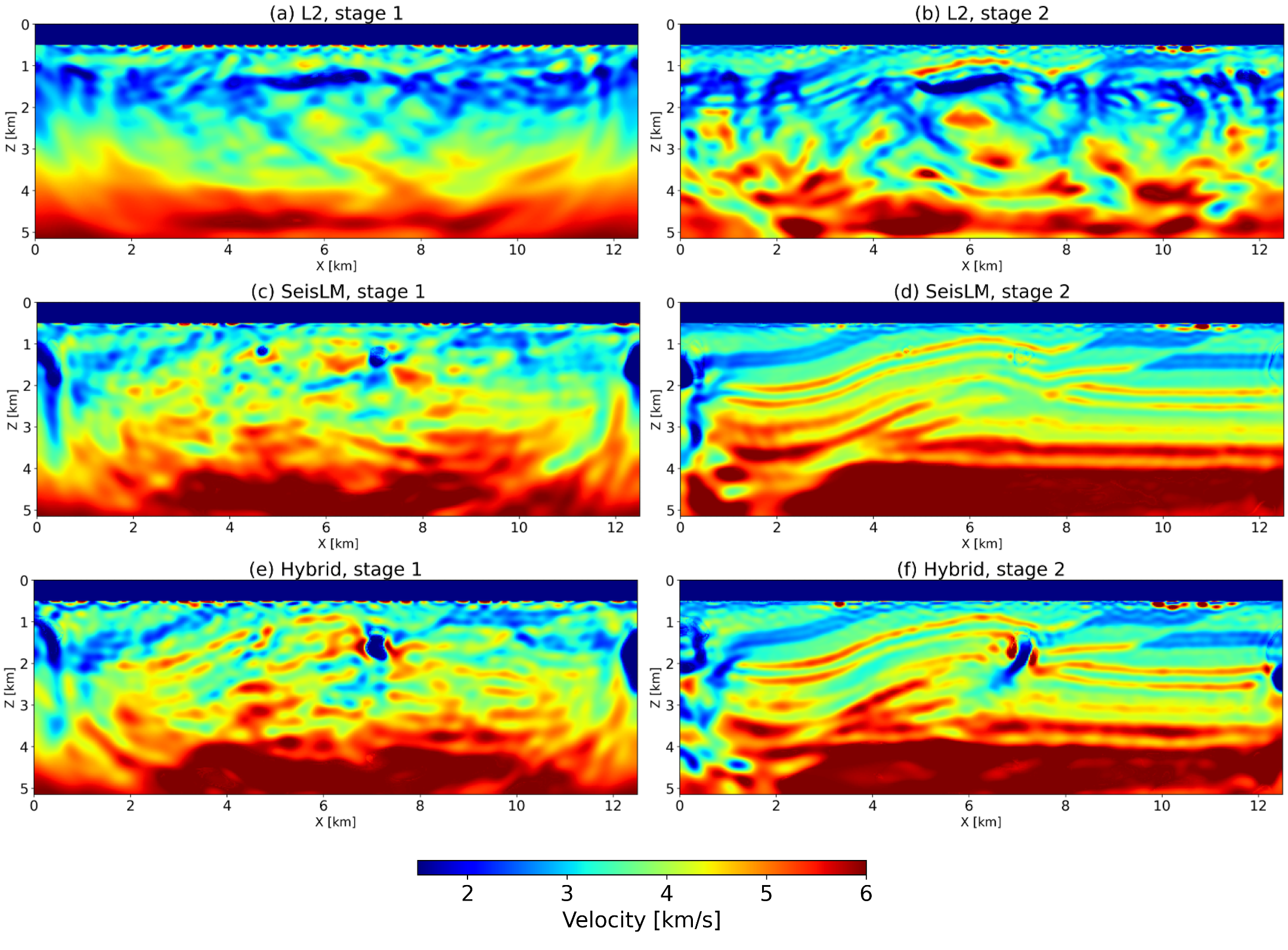}
\caption{Inverted models for the 2D Overthrust experiment. Left column: stage-1 results obtained using (a) the conventional $L_2$ objective, (c) the SeisLM feature loss, and (e) the hybrid loss. Right column: corresponding results after stage-2 $L_2$ refinement for (b) $L_2$, (d) SeisLM, and (f) the hybrid workflow.}
\label{fig:ovth_lininit_inv}
\end{figure}

Figure~\ref{fig:ovth_lininit_perf} shows the convergence histories for the 2D Overthrust experiment. The relative velocity error of the $L_2$ inversion increases almost monotonically, and the corresponding stage-1 and stage-2 models in Figures~\ref{fig:ovth_lininit_inv}(a) and (b) are dominated by cycle-skipping artifacts. In contrast, the SeisLM workflow substantially reduces the velocity error and recovers the principal layered and thrust structures, as shown in Figures~\ref{fig:ovth_lininit_inv}(c) and (d). The hybrid workflow performs less favorably in this example: its velocity error decreases less than that of the SeisLM workflow, and its reconstructed models retain pronounced artifacts near the lateral boundaries and in the central part of the model, as shown in Figures~\ref{fig:ovth_lininit_inv}(e) and (f). This degradation is consistent with the increasing $L_2$ weight reintroducing cycle-skipping sensitivity during stage 1.

\subsubsection{3D Example}
\begin{figure}[!ht]
\centering
\includegraphics[width=\textwidth]{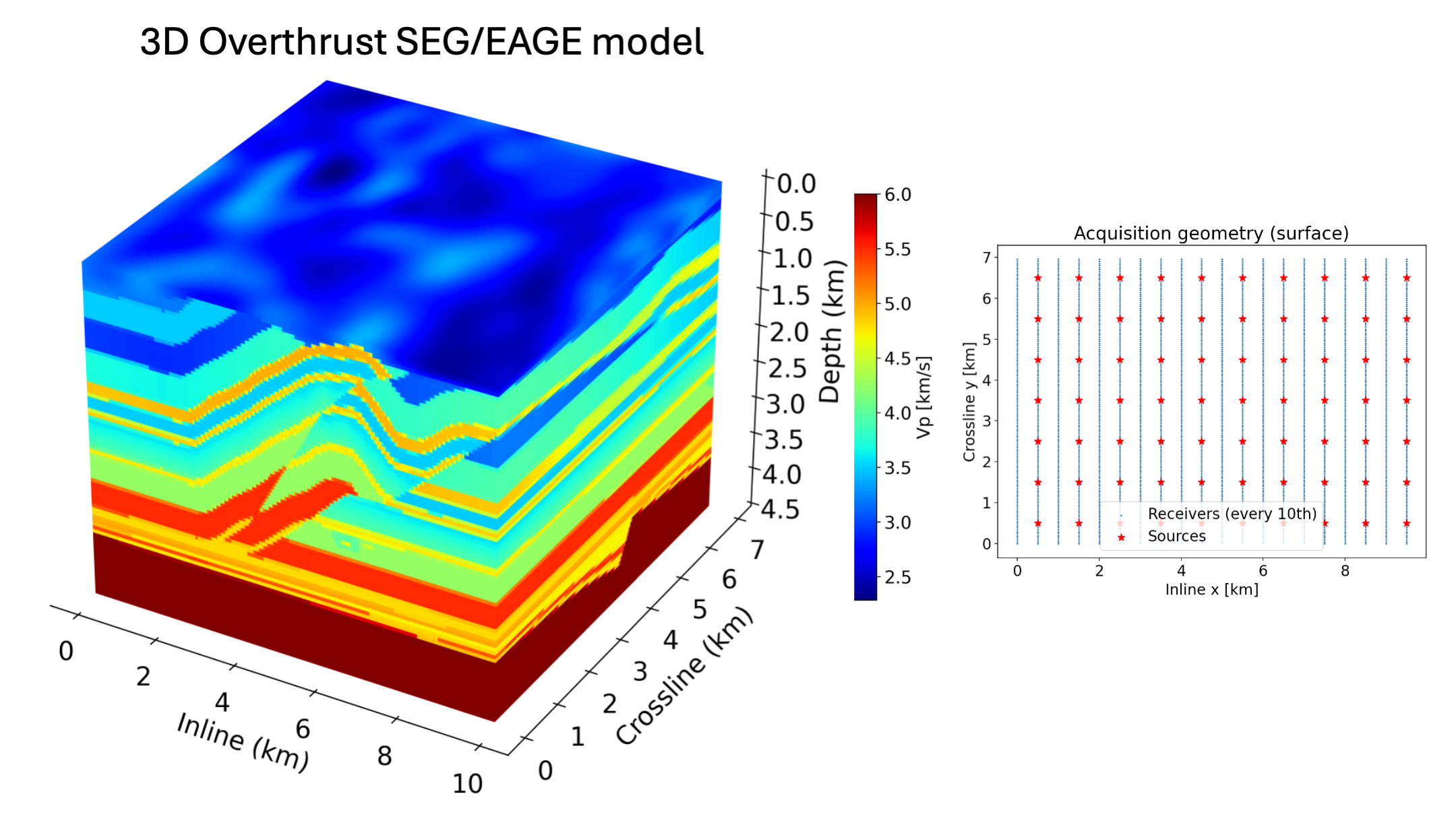}
\caption{The selected 3D Overthrust volume before adding the 500~m water layer, together with the surface acquisition footprint used for the inversion. Red stars indicate the 70 sources. Blue points indicate the receiver grid; every tenth receiver is shown for clarity.}
\label{fig:ovth_3d_model}
\end{figure}

\begin{figure}[!ht]
\centering
\includegraphics[width=\textwidth]{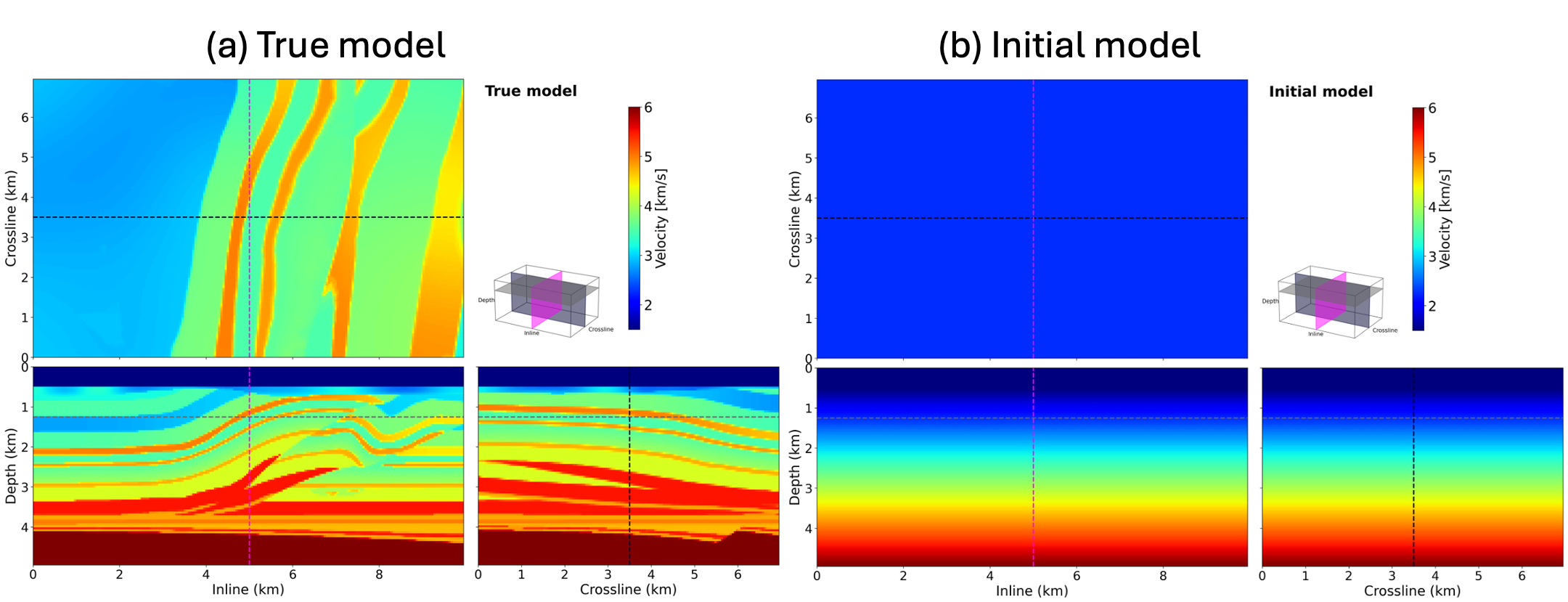}
\caption{Slices through the 3D Overthrust model: horizontal (top left), inline--depth (bottom left), and crossline--depth (bottom right). (a) True model; (b) laterally invariant linear-gradient initial model.}
\label{fig:ovth_3d_init}
\end{figure}

The final example uses a portion of the 3D Overthrust model. Starting from the original array ordered as depth, crossline, and inline, the upper 180 depth samples, first 280 crossline samples, and first 400 inline samples are selected and decimated by a factor of two in each direction. After transposition, the selected 3D volume contains $200 \times 140 \times 90$ grid points with 50~m spacing, covering approximately $10 \times 7 \times 4.5$~km before the water layer is added. After adding the 500~m water column, the working inversion model contains $200 \times 140 \times 100$ grid points and covers approximately $10 \times 7 \times 5$~km. Figure~\ref{fig:ovth_3d_model} shows the selected 3D Overthrust volume before adding the water layer, together with the surface acquisition footprint. Figure~\ref{fig:ovth_3d_init} compares representative slices through the true model and the laterally invariant linear-gradient initial model after addition of the water layer.

The survey contains 70 sources arranged on a $10 \times 7$ grid. Source spacing is 1~km in both horizontal directions, with the first source at $(x,y)=(0.5,0.5)$~km and the last at $(9.5,6.5)$~km. A dense receiver grid of $200 \times 140$ locations covers the surface with 50~m spacing, giving 28,000 receivers per shot. Sources and receivers are located at $z=0$, and each shot is recorded for 5~s. To reduce the memory required to cache the observed features for 70 shots with 28,000 traces per shot, the last-layer features are averaged over their temporal positions before the feature discrepancy is evaluated. Each trace therefore contributes one 240-dimensional feature vector rather than an $L \times 240$ feature matrix. All other preprocessing and gradient calculations follow the procedure described in Section~\ref{sec:feature_loss}. The hybrid loss is not evaluated in 3D. It produces results similar to the SeisLM loss in the Marmousi example, while performing less favorably in the 2D Overthrust experiment. We therefore restrict the computationally more demanding 3D comparison to the conventional $L_2$ and SeisLM objectives.

\begin{figure}[!ht]
\centering
\includegraphics[width=\textwidth]{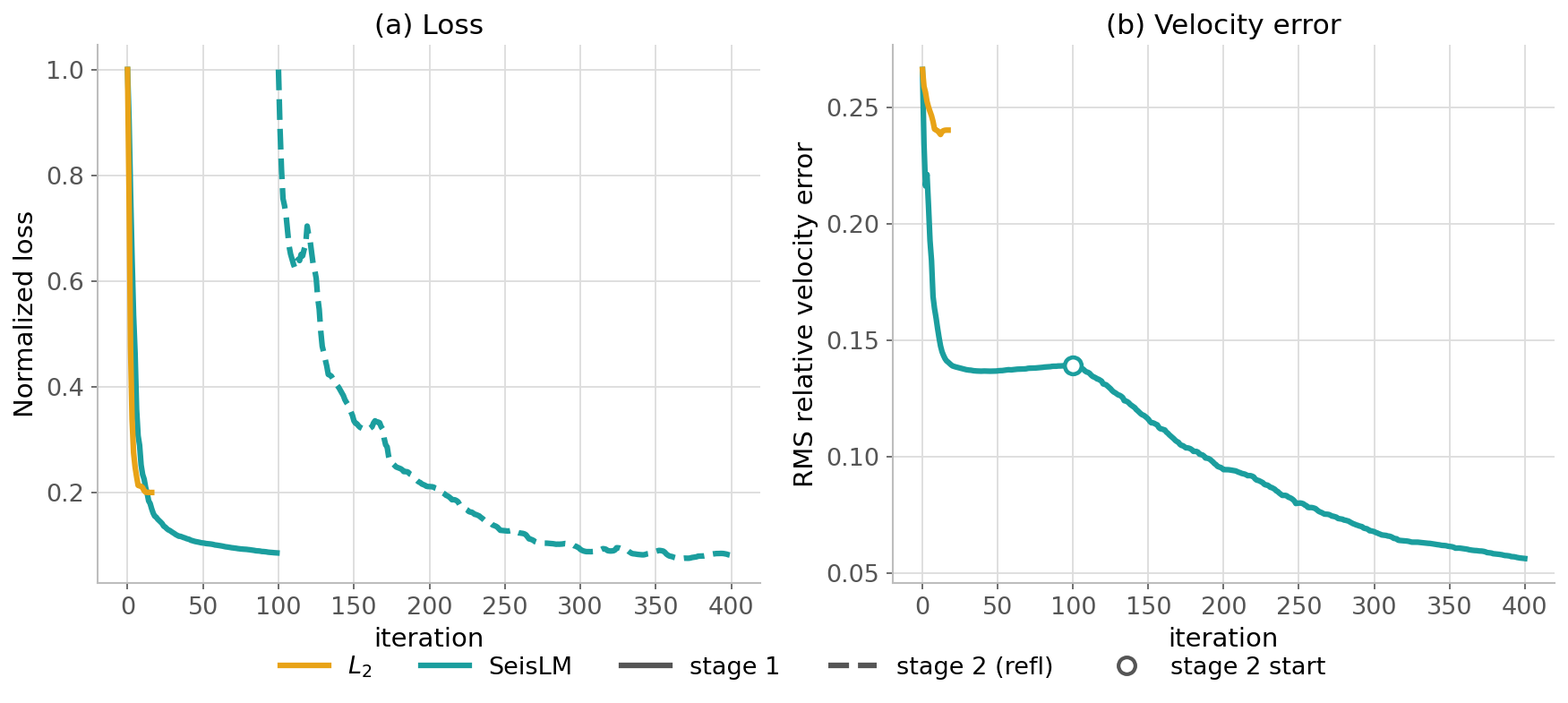}
\caption{3D Overthrust convergence: (a) stage-normalized objectives and (b) relative velocity error. The $L_2$ inversion terminates after 17 iterations; the circle marks the start of stage 2.}
\label{fig:ovth_3d_perf}
\end{figure}

As in the 2D experiment, the inversion is performed in two stages. Stage 1 suppresses reflection energy with a time--space mask and uses either the conventional $L_2$ objective or the SeisLM feature loss. Stage 2 uses the complete data set and the conventional $L_2$ objective. Figure~\ref{fig:ovth_3d_perf} shows the normalized objective values and relative velocity errors.

\begin{figure}[!ht]
\centering
\includegraphics[width=\textwidth]{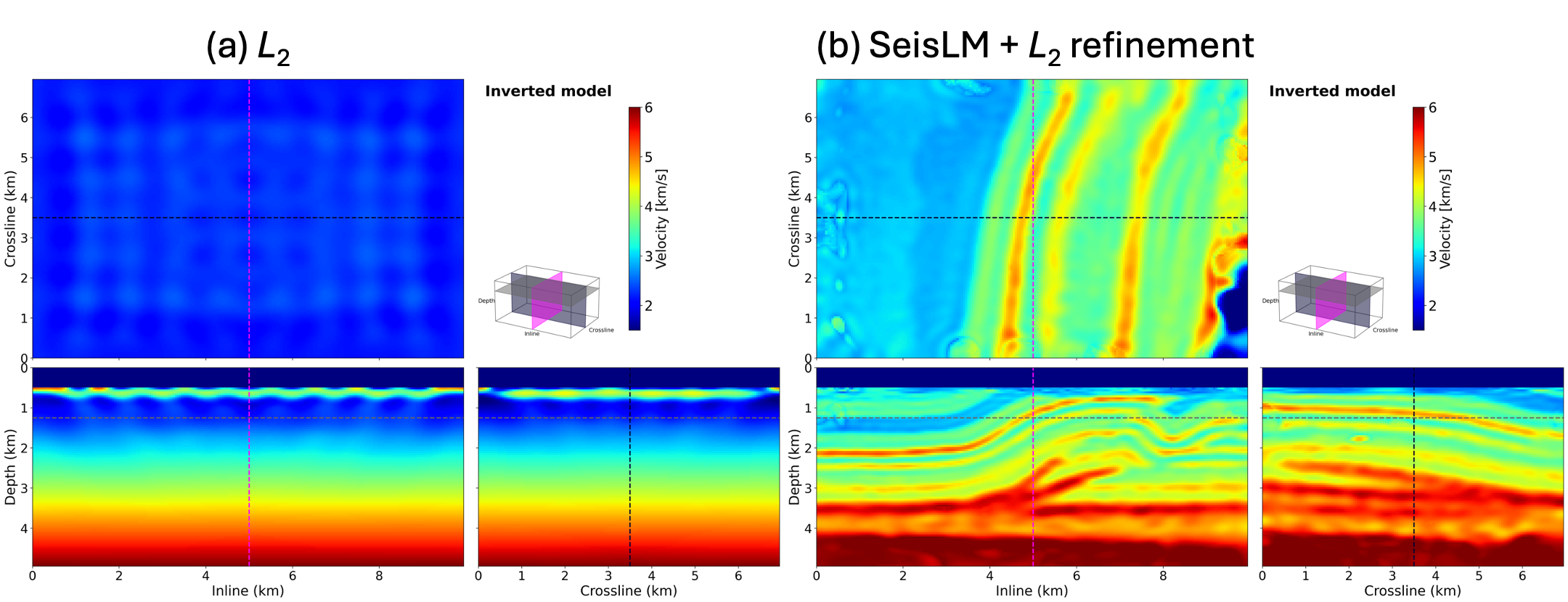}
\caption{Final results of the 3D Overthrust workflows: (a) conventional $L_2$ inversion after termination at 17 stage-1 iterations and (b) result obtained after 100 iterations of stage-1 SeisLM inversion followed by 300 iterations of stage-2 $L_2$ refinement. Dashed lines indicate the locations of the displayed inline and crossline slices.}
\label{fig:ovth_3d_inv}
\end{figure}

The $L_2$ inversion terminates after 17 stage-1 iterations when the relative reduction in the objective falls below the convergence tolerance. During these iterations, the relative velocity error decreases only from 0.266 to 0.240. Because the inversion stalls before producing a suitable background model, no stage-2 refinement is performed. The corresponding result in Figure~\ref{fig:ovth_3d_inv}(a) remains close to the initial linear gradient, apart from a thin update immediately below the water bottom, and recovers no structure at depth.

In contrast, the stage-1 SeisLM inversion reduces the relative velocity error from 0.266 to 0.139 over 100 iterations. The resulting model provides a sufficiently accurate background for stage-2 $L_2$ refinement, which reduces the error further to 0.056 over 300 iterations. The final model in Figure~\ref{fig:ovth_3d_inv}(b) recovers the layered sequence above the thrust, the dipping thrust structure, and the high-velocity basement near 4.2~km depth.

Starting from the same laterally invariant linear-gradient model, the conventional $L_2$ objective stalls without recovering the subsurface structure, whereas the SeisLM objective constructs a background model from which conventional waveform-based refinement converges. This result indicates that the benefit of the feature-space comparison extends to 3D inversion, although temporal averaging is used to reduce its memory requirements.

\section{Discussion}
This work compares modeled and observed seismic data in the feature space of a pretrained seismic foundation model. SeisLM provides waveform representations learned from a large collection of seismic recordings without task-specific training or fine-tuning for FWI. The encoder remains frozen throughout the inversion, so the only optimized parameters are those describing the velocity model. Because the encoder is differentiable with respect to its input, the derivative of the feature loss with respect to the modeled data can be used as an adjoint source without modifying the wave-equation solver.

The SeisLM and conventional $L_2$ objectives compare different representations of the data. The $L_2$ objective penalizes sample-by-sample amplitude differences and is therefore sensitive to phase errors. Even when two waveforms have similar kinematics, a shift of approximately half a cycle can produce a large residual with an incorrect gradient direction. In contrast, the SeisLM objective compares latent representations of standardized traces. The time-shift diagnostics suggest that these representations emphasize event-scale temporal structure rather than exact sample alignment, producing a smoother response to timing errors. This behavior is analogous to perceptual losses in computer vision, where distances in a pretrained feature space can provide smoother optimization landscapes than pixel-wise comparisons \cite{johnson2016perceptual, gatys2016image}. However, per-trace standardization also removes absolute amplitude information. The feature loss is therefore better suited to recovering the kinematic background model than to replacing the conventional waveform objective throughout the inversion.

The diagnostic experiments indicate that the observed time-shift robustness depends on pretraining and does not arise from network depth alone. In the toy example in Figure~\ref{fig:seislm_boa}, the randomly initialized encoder tested here provides no clear advantage over the data-space $L_2$ objective, whereas the pretrained representation produces a broader and smoother basin around the correct alignment. The layerwise test on the synthetic observed Marmousi data in Figure~\ref{fig:layerwise} shows a similar pattern. The convolutional features retain the oscillatory behavior of the data-space distance, while the pretrained transformer blocks extend the range of timing errors over which the feature distance increases monotonically. This range broadens with encoder depth, supporting the use of the last-layer representations. The randomly initialized encoder does not show the same layerwise development. These results are consistent with pretraining on diverse seismic waveforms producing representations that are less sensitive to moderate timing perturbations while retaining useful waveform structure \cite{bommasani2021opportunities, baevski2020wav2vec}. Because both diagnostics apply controlled time shifts, they explain one mechanism that may reduce cycle skipping without establishing robustness to every form of modeling or acquisition error.

Two forms of feature comparison are used in the numerical experiments. The Marmousi and 2D Overthrust examples compare the complete sequence of last-layer representations, so each trace contributes an $L \times 240$ feature matrix and retains the temporal position of each feature. To reduce the memory required to cache the observed features in the 3D experiment, the representations are averaged over the temporal dimension before comparison, reducing each trace to a 240-dimensional vector. Despite the loss of explicit temporal localization, the averaged representation produces a background model from which conventional $L_2$ refinement converges. However, the 2D and 3D results do not provide a direct comparison between the full and averaged representations. The effect of temporal averaging on the objective landscape and reconstructed model therefore remains to be evaluated.

The main value of the feature loss lies in constructing a kinematically useful model for subsequent waveform-based refinement. The quality of a stage-1 model should therefore be judged not only by its visual appearance or pointwise velocity error as well as by whether it lies within the basin of attraction of the conventional waveform objective. In the Marmousi example, the $L_2$ objective reduces the data loss while increasing the velocity error, indicating convergence toward an inaccurate background model. The SeisLM and hybrid objectives instead reduce the velocity error and provide better initial models for reflection-based refinement. Their convergence trajectories and final reconstructions are similar.

The more challenging 2D Overthrust experiment begins from a laterally invariant linear-gradient model that contains no lateral structural information. In this case, the feature-only workflow performs better than the hybrid workflow. The increasing $L_2$ contribution in the hybrid objective appears to reintroduce cycle-skipping sensitivity before the background model is sufficiently accurate. The hybrid loss is therefore not evaluated in the computationally more demanding 3D experiment. Instead, the 3D workflow uses the SeisLM feature loss to construct a background model from the same type of linear-gradient initialization and then applies conventional $L_2$ refinement. These results suggest that a fixed schedule for blending the objectives is not consistently beneficial and may need to account for the accuracy of the current model.

The feature-space objective changes only the data-comparison term, so its wave-equation modeling workload is identical to that of conventional FWI. The additional computation consists of a forward pass through the frozen SeisLM encoder for the modeled traces and a backward pass through the encoder to construct the adjoint source. The observed features are independent of the velocity model and are therefore computed once before inversion and cached. Unlike learned objectives that train or update an auxiliary network during inversion, the proposed method introduces neither additional trainable parameters nor a network-optimization loop.

Table~\ref{tab:cost} summarizes the measured computational cost. On a single NVIDIA A100-SXM4-80GB GPU, with an AMD EPYC 7713P host using 30 CPU cores for wave modeling, one full gradient evaluation using bfloat16 encoder precision takes 78~s for the 2D Marmousi example and 33.9~min for the 3D Overthrust example. The feature-loss calculation accounts for 14~s and 20.7~min of these totals, respectively. In the 3D experiment, bfloat16 reduces the full-gradient time from 67.0 to 33.9~min relative to float32. The corresponding model-error trajectories differ by less than 0.2\% over 30 iterations, indicating no measurable effect on the inversion over the tested interval. The larger 3D overhead primarily reflects the dense areal acquisition and the resulting number of traces processed by the encoder.

\begin{table}[!ht]
\centering
\caption{Per-iteration computational cost of the SeisLM feature loss, measured on a single NVIDIA A100-SXM4-80GB GPU with an AMD EPYC 7713P host using 30 CPU cores for wave modeling. The software environment uses PyTorch~2.6 and CUDA~12.4. Dimensions are given as shots $\times$ receivers $\times$ time samples.}
\label{tab:cost}
\footnotesize
\begin{threeparttable}
\begin{tabular}{@{}llccccc@{}}
\toprule
Model & Precision &
\begin{tabular}[c]{@{}c@{}}Per-shot\\feature loss\end{tabular} &
\begin{tabular}[c]{@{}c@{}}Feature loss\\per iteration\end{tabular} &
\begin{tabular}[c]{@{}c@{}}Wave modeling\\per iteration\tnote{a}\end{tabular} &
\begin{tabular}[c]{@{}c@{}}Full gradient\\per iteration\tnote{b}\end{tabular} &
Overhead\tnote{c} \\
\midrule
2D Marmousi & float32 & 1.91~s & 57~s & 64~s & 121~s & 47\% \\
($30\times300\times4107$) & bfloat16 & 0.48~s & 14~s & 64~s & 78~s & 18\% \\
\addlinespace
3D Overthrust & float32 & 46.1~s & 53.8~min & 13.2~min & 67.0~min & 80\% \\
($70\times28000\times1326$) & bfloat16 & 17.8~s & 20.7~min & 13.2~min & 33.9~min & 61\% \\
\bottomrule
\end{tabular}
\begin{tablenotes}
\footnotesize
\item[a] CPU-side wave modeling; mean of the two precision runs.
\item[b] Sum of the wave-modeling and feature-loss times.
\item[c] Feature-loss time divided by the full-gradient time.
\end{tablenotes}
\end{threeparttable}
\end{table}

Several limitations should be considered when interpreting these results. SeisLM was pretrained on three-component earthquake recordings, whereas the present experiments use single-component acoustic traces replicated across the three input channels. The FWI traces also differ from the pretraining data in frequency content, duration, and sampling interval. The numerical results demonstrate that useful transfer is possible under these conditions, although they do not determine how far the data can depart from the pretraining distribution before the representation becomes ineffective. In addition, the current synthetic experiments do not establish robustness to field noise, source-wavelet uncertainty, elastic effects, or other modeling errors. These factors should be tested before applying the method to field data.

Computational cost is another limitation. Transformer attention scales quadratically with sequence length, making forward and backward passes increasingly expensive for long traces and large surveys. Temporal averaging in the 3D experiment reduces the storage required for cached observed features, without reducing the cost of passing each trace through the encoder. Mixed precision substantially reduces this overhead, although the feature loss remains more expensive than a direct waveform comparison. The hybrid objective also requires feature-space evaluations throughout stage 1 and therefore provides no computational saving relative to the feature-only objective.

A more efficient strategy may be to use the feature loss only until the model enters the basin of attraction of the $L_2$ objective and then switch completely to waveform-based inversion. This would replace the fixed 100-iteration hybrid transition with an adaptive criterion based on the current inversion state. Possible indicators include changes in the feature loss, agreement between the feature-space and waveform gradients, or the stability of estimated arrival-time differences. Developing and validating such a criterion remains an open problem.

Another direction is to develop a compact encoder for FWI misfit computation. Such an encoder could be pretrained directly on synthetic exploration-seismic data with frequencies and acquisition characteristics representative of early-stage inversion. Alternatively, SeisLM could be distilled into a smaller student network using its pretrained representations as supervision. Either approach could reduce the computational cost and the domain mismatch between earthquake and exploration-seismic data.

\section{Conclusions}
This study introduces a feature-space misfit for FWI based on SeisLM, a pretrained seismic foundation model. Modeled and observed traces are compared using representations from the frozen encoder, and the resulting feature loss is differentiated with respect to the modeled data to construct an adjoint source. The method therefore integrates with a standard adjoint-state workflow without modifying the wave-equation solver or training an additional network during inversion. A scheduled hybrid loss combining the SeisLM and conventional $L_2$ objectives was also examined.

Time-shift diagnostics show that the pretrained SeisLM representation produces a broader and smoother basin around the correct alignment than either the data-space $L_2$ objective or the randomly initialized encoder tested here. The Marmousi experiment shows that the SeisLM and hybrid objectives reduce the velocity error during stage 1 and provide useful starting models for reflection-based $L_2$ refinement, with both workflows producing similar final results. In the 2D Overthrust experiment, which begins from a laterally invariant linear-gradient model, the SeisLM workflow recovers a more accurate background and final model than the conventional $L_2$ and hybrid workflows. The weaker hybrid result indicates that introducing the $L_2$ contribution too early can reintroduce cycle-skipping sensitivity. In the 3D Overthrust experiment, the conventional $L_2$ inversion stalls near the initial linear gradient, whereas the SeisLM objective constructs a background model from which subsequent $L_2$ refinement recovers the principal structures and substantially reduces the velocity error.

These results support using pretrained seismic representations as early-stage FWI objectives rather than as complete replacements for waveform-domain misfits. The feature loss can steer the inversion from an inaccurate starting model toward a more favorable basin of attraction, after which conventional $L_2$ refinement recovers higher-resolution structure. Future work should evaluate the method on noisy and field data, develop adaptive criteria for switching to the waveform objective, and reduce the computational cost through compact or distilled feature encoders.
\section*{Acknowledgments}
The authors thank KAUST and Shearwater GeoServices for supporting this research. For computer time, this research leveraged the resources of the Supercomputing Laboratory at KAUST in Thuwal, Saudi Arabia.
\section*{Data and Code Availability}
All data sets used in this study are publicly available. The Marmousi II model can be obtained from the University of Houston's AGL downloads page (\url{http://www.agl.uh.edu/downloads/downloads.htm}). The SEG/EAGE Salt and Overthrust models are available through the SEG Wiki page (\url{https://wiki.seg.org/wiki/SEG/EAGE_Salt_and_Overthrust_Models}). The source code supporting the findings of this study are publicly available on GitHub at: \url{https://github.com/DIG-Kaust/FM-FWI}
\bibliographystyle{unsrtnat}
\bibliography{References}

@article{tarantola1984inversion,
    author = {Tarantola, Albert},
    title = {Inversion of seismic reflection data in the acoustic approximation},
    journal = {Geophysics},
    volume = {49},
    number = {8},
    pages = {1259--1266},
    year = {1984},
    month = {08},
    issn = {0016-8033},
    doi = {10.1190/1.1441754},
}

@article{virieux2009overview,
  title={An overview of full-waveform inversion in exploration geophysics},
  author={Virieux, Jean and Operto, St{\'e}phane},
  journal={Geophysics},
  volume={74},
  number={6},
  pages={WCC1--WCC26},
  year={2009},
  publisher={Society of Exploration Geophysicists},
  doi={10.1190/1.3238367}
}

@article{luo1991wave,
  title={Wave-equation traveltime inversion},
  author={Luo, Yi and Schuster, Gerard T.},
  journal={Geophysics},
  volume={56},
  number={5},
  pages={645--653},
  year={1991},
  doi={10.1190/1.1443081},
  publisher={Society of Exploration Geophysicists}
}

@article{van2010correlation,
author = {Van Leeuwen, T. and Mulder, W. A.},
title = {A correlation-based misfit criterion for wave-equation traveltime tomography},
journal = {Geophysical Journal International},
volume = {182},
number = {3},
pages = {1383--1394},
doi = {10.1111/j.1365-246X.2010.04681.x},
year = {2010}
}

@article{warner2016adaptive,
  title={Adaptive waveform inversion: Theory},
  author={Warner, Michael and Guasch, Llu{\'\i}s},
  journal={Geophysics},
  volume={81},
  number={6},
  pages={R429--R445},
  year={2016},
  doi={10.1190/geo2015-0387.1},
  publisher={Society of Exploration Geophysicists}
}

@article{chi2014full,
title = {Full waveform inversion method using envelope objective function without low frequency data},
author = {Chi, Benxin and Dong, Liangguo and Liu, Yuzhu},
journal = {Journal of Applied Geophysics},
volume = {109},
pages = {36--46},
year = {2014},
doi = {10.1016/j.jappgeo.2014.07.010}
}

@article{wu2014seismic,
  title={Seismic envelope inversion and modulation signal model},
  author={Wu, Ru-Shan and Luo, Jingrui and Wu, Bangyu},
  journal={Geophysics},
  volume={79},
  number={3},
  pages={WA13--WA24},
  year={2014},
  publisher={Society of Exploration Geophysicists},
  doi={10.1190/geo2013-0294.1}
}

@article{Bozdag2011845,
  author={Bozda{\u{g}}, Ebru and Trampert, Jeannot and Tromp, Jeroen},
  title={Misfit functions for full waveform inversion based on instantaneous phase and envelope measurements},
  journal={Geophysical Journal International},
  volume={185},
  number={2},
  pages={845--870},
  year={2011},
  doi={10.1111/j.1365-246X.2011.04970.x}
}

@article{engquist2016optimal,
  title={Optimal transport for seismic full waveform inversion},
  author={Engquist, Bjorn and Froese, Brittany D. and Yang, Yunan},
  journal={Communications in Mathematical Sciences},
  volume={14},
  number={8},
  pages={2309--2330},
  year={2016},
  doi={10.4310/CMS.2016.v14.n8.a9}
}

@article{metivier2016measuring,
  title={Measuring the misfit between seismograms using an optimal transport distance: Application to full waveform inversion},
  author={M{\'e}tivier, Ludovic and Brossier, Romain and M{\'e}rigot, Quentin and Oudet, Edouard and Virieux, Jean},
  journal={Geophysical Journal International},
  volume={205},
  number={1},
  pages={345--377},
  year={2016},
  doi={10.1093/gji/ggw014},
  publisher={Oxford University Press}
}

@article{yang2018application,
  title={Application of optimal transport and the quadratic {Wasserstein} metric to full-waveform inversion},
  author={Yang, Yunan and Engquist, Bj{\"o}rn and Sun, Junzhe and Hamfeldt, Brittany F},
  journal={Geophysics},
  volume={83},
  number={1},
  pages={R43--R62},
  year={2018},
  doi={10.1190/geo2016-0663.1},
  publisher={Society of Exploration Geophysicists}
}

@article{ramos2019long,
  title={Long-wavelength {FWI} updates in the presence of cycle skipping},
  author={Ramos-Mart{\'\i}nez, Jaime and Qiu, Lingyun and Valenciano, Alejandro A and Jiang, Xiaoyan and Chemingui, Nizar},
  journal={The Leading Edge},
  volume={38},
  number={3},
  pages={193--196},
  year={2019},
  doi={10.1190/tle38030193.1},
  publisher={Society of Exploration Geophysicists}
}

@article{ma2013wave,
  title={Wave-equation reflection traveltime inversion with dynamic warping and full-waveform inversion},
  author={Ma, Yong and Hale, Dave},
  journal={Geophysics},
  volume={78},
  number={6},
  pages={R223--R233},
  year={2013},
  doi={10.1190/geo2013-0004.1},
  publisher={Society of Exploration Geophysicists}
}

@article{chen2022cycle,
  title={Cycle-skipping mitigation using misfit measurements based on differentiable dynamic time warping},
  author={Chen, Fuqiang and Peter, Daniel and Ravasi, Matteo},
  journal={Geophysics},
  volume={87},
  number={4},
  pages={R325--R335},
  year={2022},
  doi={10.1190/geo2021-0598.1},
  publisher={Society of Exploration Geophysicists}
}

@inproceedings{kalita2023soft,
  author    = {Kalita, Mahesh and Purcell, Chris and Casasanta, Lorenzo},
  title     = {Soft-dynamic time warping divergence as a misfit measure in full-waveform inversion},
  booktitle = {84th {EAGE} Annual Conference and Exhibition},
  volume    = {2023},
  number    = {1},
  pages     = {1--5},
  year      = {2023},
  doi       = {10.3997/2214-4609.202310489},
  publisher = {European Association of Geoscientists and Engineers}
}

@article{pladys2021cycle,
    author = {Pladys, Arnaud and Brossier, Romain and Li, Yubing and M{\'e}tivier, Ludovic},
    title = {On cycle-skipping and misfit function modification for full-wave inversion: Comparison of five recent approaches},
    journal = {Geophysics},
    volume = {86},
    number = {4},
    pages = {R563--R587},
    year = {2021},
    doi = {10.1190/geo2020-0851.1}
}

@article{vanleeuwen2013mitigating,
  title={Mitigating local minima in full-waveform inversion by expanding the search space},
  author={van Leeuwen, Tristan and Herrmann, Felix J.},
  journal={Geophysical Journal International},
  volume={195},
  number={1},
  pages={661--667},
  year={2013},
  doi={10.1093/gji/ggt258},
  publisher={Oxford University Press}
}

@article{aghamiry2019improving,
  title={Improving full-waveform inversion by wavefield reconstruction with the alternating direction method of multipliers},
  author={Aghamiry, Hossein S. and Gholami, Ali and Operto, St{\'e}phane},
  journal={Geophysics},
  volume={84},
  number={1},
  pages={R139--R162},
  year={2019},
  doi={10.1190/geo2018-0093.1},
  publisher={Society of Exploration Geophysicists}
}

@article{sirgue2004efficient,
  title={Efficient waveform inversion and imaging: A strategy for selecting temporal frequencies},
  author={Sirgue, Laurent and Pratt, R. Gerhard},
  journal={Geophysics},
  volume={69},
  number={1},
  pages={231--248},
  year={2004},
  publisher={Society of Exploration Geophysicists},
  doi={10.1190/1.1649391}
}

@inproceedings{sirgue2006importance,
  title={The importance of low frequency and large offset in waveform inversion},
  author={Sirgue, Laurent},
  booktitle={68th EAGE Conference and Exhibition incorporating SPE EUROPEC 2006},
  year={2006},
  eid={cp-2-00154},
  doi={10.3997/2214-4609.201402146},
  publisher={European Association of Geoscientists and Engineers}
}

@article{wang2019salt,
  author={Wang, Ping and Zhang, Zhigang and Mei, Jiawei and Lin, Feng and Huang, Rongxin},
  title={Full-waveform inversion for salt: A coming of age},
  journal={The Leading Edge},
  volume={38},
  number={3},
  pages={204--213},
  year={2019},
  doi={10.1190/tle38030204.1}
}

@article{cobo2019full,
  title={Full-waveform inversion for imaging and geologic interpretation: A deepwater {Gulf of Mexico} case study},
  author={Cobo, Yannick and Calder{\'o}n-Mac{\'i}as, Carlos and Chi, Shihong},
  journal={Interpretation},
  volume={7},
  number={2},
  pages={SB11--SB21},
  year={2019},
  doi={10.1190/INT-2018-0163.1}
}

@inproceedings{ji2025acquisition,
  author={Ji, Shuo and Yang, Zhiping and De Marco Centeno, Ricardo and Zhu, Huifeng and Wray, Brad},
  title={Acquisition and early imaging of a long-offset, low-frequency sparse node survey in the {Gulf of Mexico}},
  booktitle={Fifth International Meeting for Applied Geoscience and Energy Expanded Abstracts},
  pages={2920--2924},
  year={2025},
  doi={10.1190/image2025-4315295.1},
  publisher={Society of Exploration Geophysicists}
}

@ARTICLE{wu2019inversionnet,
  author={Wu, Yue and Lin, Youzuo},
  journal={IEEE Transactions on Computational Imaging}, 
  title={{InversionNet}: An efficient and accurate data-driven full waveform inversion},
  year={2020},
  volume={6},
  pages={419--433},
  doi={10.1109/TCI.2019.2956866}
  }

@article{yang2019deep,
  title={Deep-learning inversion: A next-generation seismic velocity model building method},
  author={Yang, Fangshu and Ma, Jianwei},
  journal={Geophysics},
  volume={84},
  number={4},
  pages={R583--R599},
  year={2019},
  doi={10.1190/geo2018-0249.1},
  publisher={Society of Exploration Geophysicists}
}

@inproceedings{kumar2022deep,
  author    = {Kumar, A. and Hampson, G. and Rayment, T. and Burgess, T.},
  title     = {A deep learning inverse Hessian for least-squares migration},
  booktitle = {83rd EAGE Annual Conference and Exhibition},
  volume    = {2022},
  number    = {1},
  pages     = {1--5},
  year      = {2022},
  doi       = {10.3997/2214-4609.202210158},
  publisher = {European Association of Geoscientists and Engineers}
}

@article{liu2022deep,
  title={Deep learning-based point-spread function deconvolution for migration image deblurring},
  author={Liu, Cewen and Sun, Mengyao and Dai, Nanxun and Wu, Wei and Wei, Yanwen and Guo, Mingjie and Fu, Haohuan},
  journal={Geophysics},
  volume={87},
  number={4},
  pages={S249--S265},
  year={2022},
  doi={10.1190/geo2020-0904.1},
  publisher={Society of Exploration Geophysicists}
}

@article{sun2022mlmisfit,
  author={Sun, Bingbing and Alkhalifah, Tariq},
  title={{ML}-misfit: A neural network formulation of the misfit function for full-waveform inversion},
  journal={Frontiers in Earth Science},
  volume={10},
  pages={1011825},
  year={2022},
  doi={10.3389/feart.2022.1011825}
}

@article{yang2023fwigan,
  author={Yang, Fangshu and Ma, Jianwei},
  title={{FWIGAN}: Full-Waveform Inversion via a Physics-Informed Generative Adversarial Network},
  journal={Journal of Geophysical Research: Solid Earth},
  volume={128},
  number={4},
  pages={e2022JB025493},
  year={2023},
  doi={10.1029/2022JB025493}
}

@article{wang2026fftinversiongan,
  author={Wang, Cong and Huang, Xingguo and Mousavi, S. Mostafa and Li, Yue},
  title={An unsupervised inversion framework in the frequency domain using a {W}asserstein generative adversarial network},
  journal={Geophysical Journal International},
  volume={246},
  number={2},
  pages={ggag192},
  year={2026},
  doi={10.1093/gji/ggag192}
}

@article{saad2024siamesefwi,
  title={{SiameseFWI}: A deep learning network for enhanced full waveform inversion},
  author={Saad, Omar M. and Harsuko, Randy and Alkhalifah, Tariq},
  journal={Journal of Geophysical Research: Machine Learning and Computation},
  volume={1},
  number={3},
  pages={e2024JH000227},
  year={2024},
  doi={10.1029/2024JH000227}
}

@article{saad2025f,
  title={{F-SiameseFWI}: A novel deep-learning framework for multisource full-waveform inversion},
  author={Saad, Omar M. and Alkhalifah, Tariq},
  journal={Geophysics},
  volume={90},
  number={4},
  pages={R221--R230},
  year={2025},
  doi={10.1190/geo2024-0785.1},
  publisher={Society of Exploration Geophysicists}
}

@inproceedings{devlin2019bert,
  title={{BERT}: Pre-training of deep bidirectional transformers for language understanding},
  author={Devlin, Jacob and Chang, Ming-Wei and Lee, Kenton and Toutanova, Kristina},
  booktitle={Proceedings of the 2019 Conference of the North American Chapter of the Association for Computational Linguistics: Human Language Technologies},
  volume={1},
  pages={4171--4186},
  year={2019},
  address={Minneapolis, MN},
  publisher={Association for Computational Linguistics},
  doi={10.18653/v1/N19-1423},
  url={https://aclanthology.org/N19-1423/}
}

@article{radford2019language,
  title={Language models are unsupervised multitask learners},
  author={Radford, Alec and Wu, Jeffrey and Child, Rewon and Luan, David and Amodei, Dario and Sutskever, Ilya and others},
  journal={OpenAI Blog},
  volume={1},
  number={8},
  pages={9},
  year={2019}
}

@inproceedings{baevski2020wav2vec,
  title={{wav2vec} 2.0: A framework for self-supervised learning of speech representations},
  author={Baevski, Alexei and Zhou, Henry and Mohamed, Abdelrahman and Auli, Michael},
  booktitle={Advances in Neural Information Processing Systems},
  volume={33},
  pages={12449--12460},
  year={2020},
  publisher={Curran Associates, Inc.},
  url={https://proceedings.neurips.cc/paper/2020/hash/92d1e1eb1cd6f9fba3227870bb6d7f07-Abstract.html}
}

@article{hsu2021hubert,
  title={{HuBERT}: Self-supervised speech representation learning by masked prediction of hidden units},
  author={Hsu, Wei-Ning and Bolte, Benjamin and Tsai, Yao-Hung Hubert and Lakhotia, Kushal and Salakhutdinov, Ruslan and Mohamed, Abdelrahman},
  journal={IEEE/ACM Transactions on Audio, Speech, and Language Processing},
  volume={29},
  pages={3451--3460},
  year={2021},
  doi={10.1109/TASLP.2021.3122291},
  publisher={IEEE}
}

@inproceedings{he2022masked,
  title={Masked autoencoders are scalable vision learners},
  author={He, Kaiming and Chen, Xinlei and Xie, Saining and Li, Yanghao and Doll{\'a}r, Piotr and Girshick, Ross},
  booktitle={2022 IEEE/CVF Conference on Computer Vision and Pattern Recognition (CVPR)},
  pages={15979--15988},
  year={2022},
  doi={10.1109/CVPR52688.2022.01553},
  publisher={IEEE}
}

@article{harsuko2022storseismic,
  title={{StorSeismic}: A new paradigm in deep learning for seismic processing},
  author={Harsuko, Randy and Alkhalifah, Tariq A.},
  journal={IEEE Transactions on Geoscience and Remote Sensing},
  volume={60},
  pages={1--15},
  year={2022},
  doi={10.1109/TGRS.2022.3216660},
  publisher={IEEE}
}

@article{sheng2023seismic,
  title={Seismic foundation model: A next generation deep-learning model in geophysics},
  author={Sheng, Hanlin and Wu, Xinming and Si, Xu and Li, Jintao and Zhang, Sibo and Duan, Xudong},
  journal={Geophysics},
  volume={90},
  number={2},
  pages={IM59--IM79},
  year={2025},
  doi={10.1190/geo2024-0262.1}
}

@misc{ordonez2026ncs,
  title={{NCS}-Models: Seismic foundation models trained on the {Norwegian} repository of public data},
  author={Ordonez, Alba and Forgaard, Theodor Johannes Line and Wade, David and Bugge, Aina Juell and Nese, Hakon and Waldeland, Anders Ueland},
  year={2026},
  eprint={2603.23211},
  archivePrefix={arXiv},
  primaryClass={physics.geo-ph},
  doi={10.48550/arXiv.2603.23211}
}

@article{si2024seisclip,
  title={{SeisCLIP}: A seismology foundation model pre-trained by multimodal data for multipurpose seismic feature extraction},
  author={Si, Xu and Wu, Xinming and Sheng, Hanlin and Zhu, Jun and Li, Zefeng},
  journal={IEEE Transactions on Geoscience and Remote Sensing},
  volume={62},
  pages={1--13},
  year={2024},
  doi={10.1109/TGRS.2024.3354456},
  publisher={IEEE}
}

@article{li2024seist,
  author={Li, Sen and Yang, Xu and Cao, Anye and Wang, Changbin and Liu, Yaoqi and Liu, Yapeng and Niu, Qiang},
  title={{SeisT}: A foundational deep-learning model for earthquake monitoring tasks},
  journal={IEEE Transactions on Geoscience and Remote Sensing},
  volume={62},
  pages={1--15},
  year={2024},
  doi={10.1109/TGRS.2024.3371503}
}

@article{liu2024seislm,
  title={{SeisLM}: A foundation model for seismic waveforms},
  author={Liu, Tianlin and M{\"u}nchmeyer, Jannes and Laurenti, Laura and Marone, Chris and de Hoop, Maarten V. and Dokmani{\'c}, Ivan},
  journal={arXiv preprint arXiv:2410.15765},
  year={2024},
  doi={10.48550/arXiv.2410.15765}
}

@inproceedings{johnson2016perceptual,
  title={Perceptual losses for real-time style transfer and super-resolution},
  author={Johnson, Justin and Alahi, Alexandre and Fei-Fei, Li},
  booktitle={Computer Vision -- ECCV 2016},
  pages={694--711},
  year={2016},
  publisher={Springer International Publishing},
  address={Cham},
  doi={10.1007/978-3-319-46475-6_43}
}

@inproceedings{gatys2016image,
  title={Image style transfer using convolutional neural networks},
  author={Gatys, Leon A. and Ecker, Alexander S. and Bethge, Matthias},
  booktitle={2016 IEEE Conference on Computer Vision and Pattern Recognition (CVPR)},
  pages={2414--2423},
  year={2016},
  doi={10.1109/CVPR.2016.265},
  publisher={IEEE}
}

@inproceedings{dosovitskiy2016generating,
  title={Generating images with perceptual similarity metrics based on deep networks},
  author={Dosovitskiy, Alexey and Brox, Thomas},
  booktitle={Advances in Neural Information Processing Systems},
  volume={29},
  pages={658--666},
  year={2016},
  publisher={Curran Associates, Inc.},
  url={https://proceedings.neurips.cc/paper/2016/hash/371bce7dc83817b7893bcdeed13799b5-Abstract.html}
}

@inproceedings{simonyan2015very,
  title={Very deep convolutional networks for large-scale image recognition},
  author={Simonyan, Karen and Zisserman, Andrew},
  booktitle={International Conference on Learning Representations},
  year={2015},
  doi={10.48550/arXiv.1409.1556},
  url={https://arxiv.org/abs/1409.1556}
}

@article{woollam2022seisbench,
  title={{SeisBench}: A toolbox for machine learning in seismology},
  author={Woollam, Jack and M{\"u}nchmeyer, Jannes and Tilmann, Frederik and Rietbrock, Andreas and Lange, Dietrich and Bornstein, Thomas and Diehl, Tobias and Giunchi, Carlo and Haslinger, Florian and Jozinovi{\'c}, Dario and Michelini, Alberto and Saul, Joachim and Soto, Hugo},
  journal={Seismological Research Letters},
  volume={93},
  number={3},
  pages={1695--1709},
  year={2022},
  doi={10.1785/0220210324}
}

@article{liu1989limited,
  title={On the limited memory {BFGS} method for large scale optimization},
  author={Liu, Dong C. and Nocedal, Jorge},
  journal={Mathematical Programming},
  volume={45},
  number={1--3},
  pages={503--528},
  year={1989},
  doi={10.1007/BF01589116},
  publisher={Springer}
}

@inproceedings{brougois1990marmousi,
  author    = {Brougois, A. and Bourget, M. and Lailly, P. and Poulet, M. and Ricarte, P. and Versteeg, R.},
  title     = {Marmousi, model and data},
  booktitle = {Proceedings of the 1990 {EAEG} Workshop on Practical Aspects of Seismic Data Inversion},
  year      = {1990},
  eid       = {cp-108-00002},
  doi       = {10.3997/2214-4609.201411190},
  publisher = {European Association of Geoscientists and Engineers}
}

@book{aminzadeh1997three,
  author={Aminzadeh, F. and Brac, J. and Kunz, T.},
  title={{SEG/EAGE} {3D} Salt and Overthrust Models},
  series={{SEG/EAGE} {3D} Modeling Series},
  volume={1},
  publisher={SEG/EAGE},
  address={Tulsa, OK},
  year={1997},
  note={Distribution CD of Salt and Overthrust models}
}

@article{bommasani2021opportunities,
  title={On the opportunities and risks of foundation models},
  author={Bommasani, Rishi and Hudson, Drew A. and Adeli, Ehsan and Altman, Russ and Arora, Simran and von Arx, Sydney and Bernstein, Michael S. and Bohg, Jeannette and Bosselut, Antoine and Brunskill, Emma and others},
  journal={arXiv preprint arXiv:2108.07258},
  year={2021},
  doi={10.48550/arXiv.2108.07258}
}

@article{devito-compiler,
  author={Luporini, Fabio and Louboutin, Mathias and Lange, Michael and Kukreja, Navjot and Witte, Philipp and H\"{u}ckelheim, Jan and Yount, Charles and Kelly, Paul H. J. and Herrmann, Felix J. and Gorman, Gerard J.},
  title={Architecture and performance of {Devito}, a system for automated stencil computation},
  journal={ACM Transactions on Mathematical Software},
  volume={46},
  number={1},
  articleno={6},
  numpages={28},
  year={2020},
  doi={10.1145/3374916},
  publisher={Association for Computing Machinery}
}

@article{devito-api,
  author={Louboutin, Mathias and Lange, Michael and Luporini, Fabio and Kukreja, Navjot and Witte, Philipp A. and Herrmann, Felix J. and Velesko, Paulius and Gorman, Gerard J.},
  title={{Devito} (v3.1.0): An embedded domain-specific language for finite differences and geophysical exploration},
  journal={Geoscientific Model Development},
  volume={12},
  number={3},
  pages={1165--1187},
  year={2019},
  doi={10.5194/gmd-12-1165-2019}
}

@article{2020SciPy-NMeth,
  author  = {Virtanen, Pauli and Gommers, Ralf and Oliphant, Travis E. and
            Haberland, Matt and Reddy, Tyler and Cournapeau, David and
            Burovski, Evgeni and Peterson, Pearu and Weckesser, Warren and
            Bright, Jonathan and {van der Walt}, St{\'e}fan J. and
            Brett, Matthew and Wilson, Joshua and Millman, K. Jarrod and
            Mayorov, Nikolay and Nelson, Andrew R. J. and Jones, Eric and
            Kern, Robert and Larson, Eric and Carey, C J and
            Polat, {\.I}lhan and Feng, Yu and Moore, Eric W. and
            {VanderPlas}, Jake and Laxalde, Denis and Perktold, Josef and
            Cimrman, Robert and Henriksen, Ian and Quintero, E. A. and
            Harris, Charles R. and Archibald, Anne M. and
            Ribeiro, Ant{\^o}nio H. and Pedregosa, Fabian and
            {van Mulbregt}, Paul and {SciPy 1.0 Contributors}},
  title   = {{{SciPy} 1.0: Fundamental Algorithms for Scientific
            Computing in Python}},
  journal = {Nature Methods},
  year    = {2020},
  volume  = {17},
  pages   = {261--272},
  adsurl  = {https://rdcu.be/b08Wh},
  doi     = {10.1038/s41592-019-0686-2},
}

\end{document}